# Brain scaling and diversity in musteloids

Margot MICHAUD[1,2] ; Arianna Rose HARRINGTON[3] ; Sandrine LADEVEZE[4] ; Eloy GÁLVEZ-LÓPEZ[5,6] ; Valentin FISCHER[2] ; Apolline ALFSEN[7]; Ornella C. BERTRAND[8,9]

1. Institut Polytechnique UniLaSalle, IDEALISS ULR 7519, Université d'Artois; Mont-Saint-Aignan, 76130, France.
2. Evolution & Diversity Dynamics Lab, Université de Liège, Liège, Belgium
3. Department of Biology, Southern Utah University, Cedar City, Utah, USA
4. Museum national d'Histoire naturelle, Centre de Recherche en Paleontologie, Paris, UMR 7207 CR2P MNHN/CNRS/UPMC, Sorbonne
5. PalaeoHub, Department of Archaeology, University of York, Wentworth Way, Heslington, YO10 5DD, United Kingdom
6. Centre for Integrative Anatomy, Cell and Developmental Biology, University College London, Gower Street, London, WC1E 6BT, United Kingdom
7.Evolutionary Morphology, Museum für Naturkunde – Leibniz-Institut für Evolutions- und Biodiversitätsforschung, Berlin, Germany
8. Institut Català de Paleontologia Miquel Crusafont (ICP-CERCA), Universitat Autònoma de Barcelona, Barcelona, Spain
9. Section of Mammals, Carnegie Museum of Natural History, Pittsburgh, Pennsylvania, USA

**CORRESPONDENCE** Margot Michaud, Institut Polytechnique UniLaSalle, IDEALISS ULR 7519, Université d'Artois; Mont-Saint-Aignan, 76130, France. Email : m.mar33@hotmail.fr

**PUBLISHED VERSION (VERSION OF RECORD)**

Michaud, M., Harrington, A.R., Ladeveze, S., Gálvez-López, E., Fischer, V., Alfsen, A. & Bertrand, O. (2026) Brain scaling and diversity in musteloids. *Journal of Anatomy*, 00, 1–24.

**https://doi.org/10.1111/joa.70226**

Received 12 December 2025 · Revised 25 July 2026 · Accepted 6 August 2026 · © 2026 Anatomical Society

**SHORT RUNNING TITLE** Ecological shift of brain scaling in musteloids

**KEYWORDS** Brain, endocast, Carnivora, otters, allometry, geometric morphometrics, aquatic adaptation

**FUNDING** This work was supported by the Fyssen Foundation through a postdoctoral fellowship awarded to MM, during which part of the specimen sampling was carried out. Scanning of the specimens was supported by NSF BCS 1825129 awarded to ARH. OC.B was funded by the Direcció General de Recerca de la Generalitat de Catalunya, managed by the Agència de Gestió d'Ajuts Universitaris i de Recerca, Beatriu de Pinós Programme, expedient number: 2021 BP 00042; Ministerio de Ciencia, Innovación y Universidades y Agencia Estatal de Investigación, Ayuda RYC2023-042630-I financiada por MICIU/ AEI /10.13039/501100011033 y por el FSE+. OC.B was also supported by the Generalitat de Catalunya/CERCA Programme, and the Agència de Gestió d'Ajuts Universitaris i de Recerca, Generalitat de Catalunya: 2021 SGR 00620 to David M. Alba (Institute Catala De Paleontologia).

# ABSTRACT

The diversity of brain size and organization across mammals reflects how evolutionary and ecological pressures have shaped neural architecture through time. Yet, how these selective pressures translated into measurable variation in the proportions and overall shape of the brain across species remains poorly understood. In this study, we used high-resolution virtual endocasts from 41 musteloid species (e.g. otters, skunks, raccoons) to quantify variation in endocranial composition and shape. We analyzed both the complete endocast and its principal subdivisions (olfactory bulbs, cerebrum, and hindbrain) using a combination of compositional data analysis and 3D geometric morphometrics. Phylogenetically informed regressions (PGLS) and model comparisons were used to assess the effects of endocranial size, locomotion, diet, and taxonomy on both the regional proportions and the overall shape of the endocranial cast. Musteloid brain organization is mostly driven by size-dependent scaling: larger brains have enlarged cerebra, reduced olfactory and hindbrain regions, and rounder endocasts. However, some lineages markedly deviate from this largely conserved allometry: otters (Lutrinae) exhibit proportionally enlarged cerebra and strongly reduced olfactory bulbs, whereas skunks (Mephitidae) show the reverse pattern, emphasizing enhanced olfactory investment. Feeding ecology also contributes to variation in brain organization, as piscivores show disproportionately smaller olfactory bulbs compared with other feeding guilds. In contrast, locomotor mode has little explanatory power for either endocranial composition or shape. Brain evolution in musteloids is therefore not uniform: a common scaling pattern dominates, but lineage- and diet-specific adjustments modify how brain regions are proportioned.

# INTRODUCTION

In recent years, mammalian neuroanatomy has been a fast-advancing study domain, focused on elucidating how evolutionary changes in brain size, structure, and organization have influenced sensory, motor, and cognitive diversity (Bertrand & Krubitzer, 2025). For decades, studies of brain evolution focused on encephalization, defined as the increase in brain size in relation to body mass (Jerison, 1973). While relative brain size is a key characteristic, current research increasingly underscores the importance of examining both brain proportions and shape as relevant evidence of functional evolution and ecological adaptation. A rising body of comparative studies on vertebrates indicates that differences in brain shape are substantially associated with sensory and motor adaptations across taxa (Rowe et al. 2011; Chambers et al., 2021; Lyras et al., 2023). Furthermore, changes in the volumes of individual brain regions inevitably affect the overall geometry of the endocast, since the expansion or reduction of certain areas modifies the location and curvature of neighboring structures (Bastir et al., 2011; Zollikofer and Ponce de León, 2013).

Recent advances in virtual reconstruction techniques have significantly expanded the possibilities for studying endocranial morphology. Although physical endocasts have long been used in paleontology to investigate brain morphology in fossils (Radinsky, 1977; Lyras et al., 2012; Bertrand et al., 2025), high-resolution X-ray and micro-computed tomography (μCT) now facilitate intricate 3D reconstructions of endocranial casts (i.e., three-dimensional casts of the neurocranial cavity, also simply referred as endocasts) for both extant and fossil taxa, circumventing the need for fresh brain tissue or natural cast (Pérez-Ramos & Figueirido, 2020; Zarkogiannis et al., 2024). The morphology of endocasts matches that of the brain's external surface in mammals, supporting the use of these virtual reconstructions as reliable proxies for large-scale comparative neuroanatomy studies (Dumoncel, et al., 2023; Kubicka et al., 2025). Thereby, virtual endocasts offer a reliable three-dimensional representation for measuring size and shape, especially through 3D geometric morphometrics. Although limited to the external anatomy of the brain, the use of endocasts remains particularly valuable for exploring evolu-tionary patterns, as it not only expands sample sizes but also integrates fossil species, therefore providing a broader perspective on how ecological and developmental factors have influenced brain evolution through time (Bruner et al., 2018; Bertrand & Krubitzer, 2025; Jerison et al., 2025).

Carnivora is a particularly appropriate model for examining how mammalian brains change over wide ecological gradients. The order combines a well-resolved phylogeny with an impressive range of ecological specialization, ranging from arboreal to aquatic habitats, and demonstrates exceptional diversity in their diet, social structure, and body mass (Nowak, 2005; Wilson & Mittermeier, 2009). In this paper, we focus our efforts on extant Musteloidea, including Ailuridae, Mustelidae, Mephitidae, and Procyonidae. This superfamily covers the ecological transitions most relevant to our study, featuring semi-aquatic (otters), semi-fossorial (badgers and skunks), arboreal (kinkajous), and terrestrial species within a singular closely related radiation (Macdonald et al., 2017). This targeted sampling allows us to examine whether ecological specialization results in distinctive patterns in brain composition and endocranial morphology, while providing an adequate phylogenetic framework for rigorous comparative analysis.

Moreover, previous studies on this group strongly suggest a correlation between ecology and brain evolution (Bertrand & Krubitzer, 2025). It is already well established that the skull, which houses the brain and associated sensory structures, is influenced by both locomotor and dietary ecology (Dumont et al., 2016; Law et al., 2018a). Furthermore, correlations between ecological specialization and brain evolution have been documented across mustelid species, supported by findings related to both relative brain size (Jerison, 1973; Gittleman, 1986; Finarelli & Flynn, 2007; Sakai et al., 2011; Arsznov & Sakai, 2013; Lynch & Allen, 2022) and cerebral morphology (Radinsky, 1971; Kamiya & Pirlot, 1988; Ahrens, 2014; Boch et al., 2024). Ahrens (2014) identified a strong phylogenetic signal in endocranial morphology, while also showing that arboreal procyonids exhibit expansions in brain regions associated with motor coordination and balance, particularly the flocculus and paraflocculus, as well as expansions of the frontal and parietal lobes linked more broadly to motor control. Additionally, neocortical sulcal morphology in musteloids shows variation reflecting both allometric effects and ecological or behavioral adaptations (Boch et al., 2024).

The present study aims to evaluate the impact of ecological factors on the brain evolution of Musteloidea, using endocasts as evidence of external brain morphology and composition. To address this question, we combine two complementary approaches: (1) analysis of brain composition, based on regional volume proportions (olfactory bulbs, cerebrum, hindbrain) and cortical surface areas (neocortex, paleocortex), and (2) 3D geometric morphometrics to characterize the shape of the whole endocranium and of each region separately. Brain composition reflects the functional integration (i.e., covariation) among regions through their relative volumetric proportions, whereas endocranial shape encapsulates the spatial organization of these regions within the cranial cavity.. (Javier-Torrent et al., 2021; Weisbecker et al., 2021). Because the brain nearly fills the endocranial space in mammals, both volumetric and shape data partly reflect spatial trade-offs among regions, although they emphasize different and complementary aspects of brain morphology. By combining these datasets within a phylogenetically informed framework, we aim to assess the relative contributions of size, and ecological adaptations to endocranial variations while accounting for shared evolutionary history. This approach therefore enables us to assess whether specific feeding and locomotor ecologies are associated with distinct brain proportions or evolutionary trajectories.

Because the organization of the mammalian brain reflects shared ancestry and conserved developmental processes, we expect both endocranial composition and morphology to exhibit a significant phylogenetic signal (Finlay et al., 2001; Weisbecker et al., 2021). Beyond these lineage-associated patterns, we expect dietary ecology to have a noticeable impact on brain evolution, particularly considering the strong correlation between feeding adaptations and cranial morphology (Dumont et al., 2016; Law et al., 2018a). Otters (which are members of the subfamily Lutrinae) are expected to present smaller olfactory bulbs than other groups, given that olfaction plays a limited role in underwater environments (Gittleman, 1991; Pihlström, 2008; Sulser & MacPhee, 2023; Stenwall et al., 2025). Conversely, semi-fossorial species such as skunks are anticipated to exhibit comparatively larger olfactory bulbs, indicative of their reliance on olfaction (Ribeiro et al., 2014; Onyono et al., 2017). Furthermore, given that the olfactory bulbs are among the most morphologically diverse parts of the mammalian brain and show some of the most notable evolutionary rates (Lang et al., 2025; Lopez-Aguirre et al., 2025), we anticipate that they will show the greatest degree of variability among species. Lastly, differences in the hindbrain (including the cerebellum and brainstem), which regulate balance and coordination (Surgent et al., 2019; Lacroix-Ouellette & Dubuc, 2023; Novello et al., 2023), may be associated with several locomotor behaviors. For example, species that require greater postural stability, such as semi-aquatic taxa or some arboreal species specialized for rapid locomotion in complex three-dimensional environments, tend to exhibit different hindbrain sizes and/or shapes compared to terrestrial species (Macrì et al., 2019; Sultan & Glickstein, 2007), particularly in the petrosal lobule, which seems to plays a key role in balance (Rambold et al., 2002; Bertrand et al., 2021; Lang et al., 2022; but see Ferreira-Cardoso et al., 2017).

# MATERIAL & METHODS

## Material

Our sample consisted of 113 endocasts from adult individuals representing 41 species (1 Ailuridae, 31 Mustelidae including 9 Lutrinae, 3 Mephitidae, and 6 Procyonidae) (**Figure 1**). This corresponds to 91.4% of musteloid genera, as we could not acquire any specimens for *Bassaricyon*, *Conepatus* or *Mydaus*. The number of specimens for each species ranged from one to five individuals. We selected both males and females when possible. Specimens of different provenances were included (captive, wild, or undocumented) to maximize taxonomic coverage, although wild individuals were preferentially selected when available. However, given the macroevolutionary scale of our analyses, potential intraspecific variation associated with specimen provenance is expected to be small relative to interspecific differences and therefore unlikely to influence the patterns analyzed (Revell & Reynolds, 2012; Siefert et al., 2015). Specimen information and other metadata are detailed in **Supporting Information 1.** To account for phylogenetic non-independence, we used the phylogenetic tree of the order Carnivora provided by Hassanin et al. (2021; Appendix S5), which we modified to incorporate missing species and to update several topologies following recent studies. Full details of the phylogeny used are provided in **Supporting Information 2**.

All comparative analyses were conducted at the species level (n = 41). When multiple specimens were available for a given species, measurements were first computed for each individual and then averaged per species prior to phylogenetic analyses. Using species means might under-represent intraspecific variation which could reduce statistical power for detecting ecological effects (Ives et al., 2007). This intraspecific variation may also include differences between captive and wild individuals, which can be substantial in some species (McPhee, 2004; Zuccarelli, 2004). Nevertheless, interspecific differences are expected to outweigh within-species variation and thus drive the patterns analyzed (Revell & Reynolds, 2012).

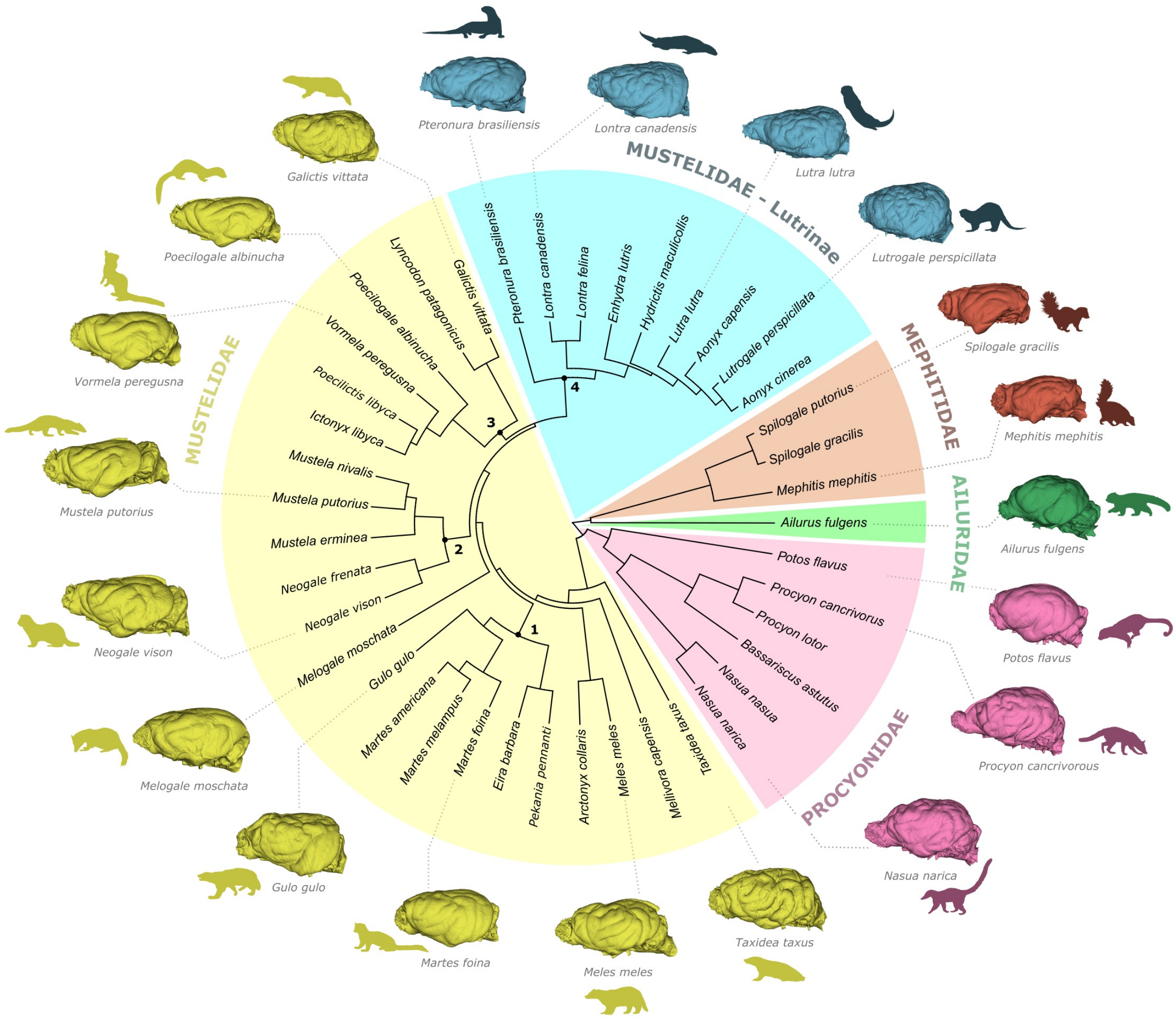

**Figure 1. Phylogeny of the four families (i.e., Ailuridae, Mephitidae, Mustelidae and Procyonidae) within Musteloidea; with virtual endocasts of representative species.** The phylogenetic relationships of the species used in this study are modified from Hassanin et al. (2021). Several nodes delineate subfamilies within the Mustelidae: 1. Guloninae; 2. Mustelinae; 3. Ictonychinae; and 4. Lutrinae. Virtual endocasts are shown for comparative purposes only and are not to scale. All the silhouettes were drawn (MM) and are available on the Phylopic website (http://www.phylopic.org/).

## Ecological and taxonomic traits

Because brain region proportions offer insights into the functional allocation of neural tissue that may not be captured by size alone, we jointly analyzed compositional variation and regional shape in relation to several ecological and taxonomic factors. Our analytical framework integrates two complementary datasets, each addressing a distinct aspect of regional endocranial variation: (1) endocranial composition, i.e., the relative volume and proportion of each major brain region describing how total brain volume is partitioned among anatomical and functional units; and (2) the global and regional shape, captured by the 3D Procrustes-aligned landmark configurations of each brain subdivision.

Ecological and behavioral information was compiled primarily from the PanTHERIA database (Jones et al., 2009) and the Handbook of the Mammals of the World (Wilson & Mittermeier, 2009). Dietary preferences were grouped into five categories (carnivorous, invertivorous – i.e., animals that eat invertebrates –, omnivorous, piscivorous and herbivorous – i.e., animals that eat fruits, leaves and other vegetable materials) based on the main food type constituting a significant portion of the species' diet, regardless of seasonal or geographic variation (Goillot et al., 2009). Locomotion categories were extracted from Gálvez-López (2021) and divided into seven modes (semi-aquatic, arboreal, semi-ar-boreal, semi-fossorial, scansorial, terrestrial and all-terrain – i.e., primarily terrestrial species that nevertheless climb well, swim efficiently, and occasionally modify burrows). Body mass data for each species were obtained from Law et al. (2018b). All classifications and body mass information are detailed in **Supporting Information 1.**

Both the brain composition and shape data were analyzed in relation to key ecological and taxonomic variables, including primary locomotion mode (referred to as 'Locomotion'), dietary category (referred to as 'Diet'), and taxonomic group. Although phylogenetic structure was already accounted for in our analyses, taxonomic group was also tested as a categorical predictor to evaluate whether families displayed distinct size-related trajectories not captured by phylogenetic covariance alone. Given the considerable morphological and ecological distinctiveness within the Mustelidae family (Dumont et al., 2015; Law et al., 2018a; Michaud et al., 2018), we implemented two taxonomic coding schemes: (1) 'Family', following classical taxonomy at family level, and (2) 'Subfamily', in which we subdivided species into Guloninae, Helictidinae, Ictonychinae, Lutrinae, Mustelinae, and the badgers (genera *Arctonyx*, *Meles, Mellivora*, and *Taxidea*), which occupy a basal position in the mustelid tree and form a paraphyletic assemblage. To ensure sufficient statistical power, Guloninae and Helictidinae were merged into a single category, and all badger genera were similarly grouped together.

## Endocast segmentation

Crania were digitized using X-ray and micro-computed tomography scans or downloaded from public repositories (e.g. Morphosource, Digimorph), with a resolution ranging from 25 µm to 1 mm (See **Supporting Information 3** for details). Three-dimensional virtual endocasts were then extracted manually following a threshold-based segmentation procedure (Balanoff et al., 2016) under AVIZO V. 8.1 (VSG SAS, Bordeaux). The channels containing the cranial nerves and the vessels exiting the endocranial cavity were closed manually following the recommendations provided by Czeibert et al. (2020) based on carnivorans. In addition, closure of the foramen magnum was standardized by drawing a straight plane between the basion, the opisthion and the lateral margins of the foramen magnum. The results of the 3D endocast segmentation of each specimen were checked before exporting them for further analyses.

## 3D Geometric Morphometrics

We used both anatomical and sliding landmarks to encompass the complex shape of the endocasts (**Figure 2**, **Table 1**). A total of 47 landmarks, 116 semilandmarks along 20 curves, and 96 surface semilandmarks were placed on each extracted endocast surface using Stratovan Checkpoint (Stratovan, Davis, CA, USA). The set of fixed landmarks was distributed across the main brain regions, with 12 on the olfactory bulbs, 16 on the cerebrum, and 19 on the hindbrain. The landmarks were chosen to best describe shape variations of the overall endocast and the different structures that compose it, including the olfactory bulbs, the cerebrum, and the hindbrain (corresponding to the cerebellum, the pons and the medulla oblongata). Regional partitioning of the endocast was performed according to Flink & Werdelin (2022) (**Figure 3**).

Landmark subsets were created to further analyze shape variation on the olfactory bulbs (OB), cerebrum (CR), and hindbrain (HB) shape variations. Some landmarks and curves used in our analyses were based on previous studies investigating mammalian endocast morphology (Ahrens, 2014; Bertrand et al., 2019). In addition, surface semilandmarks were placed on the surface of the cerebrum because this part of the endocast lacks clear anatomical landmarks. Semilandmarks were allowed to slide along the curves and surfaces to minimize the bending energy (Gunz & Mitteroecker, 2013).

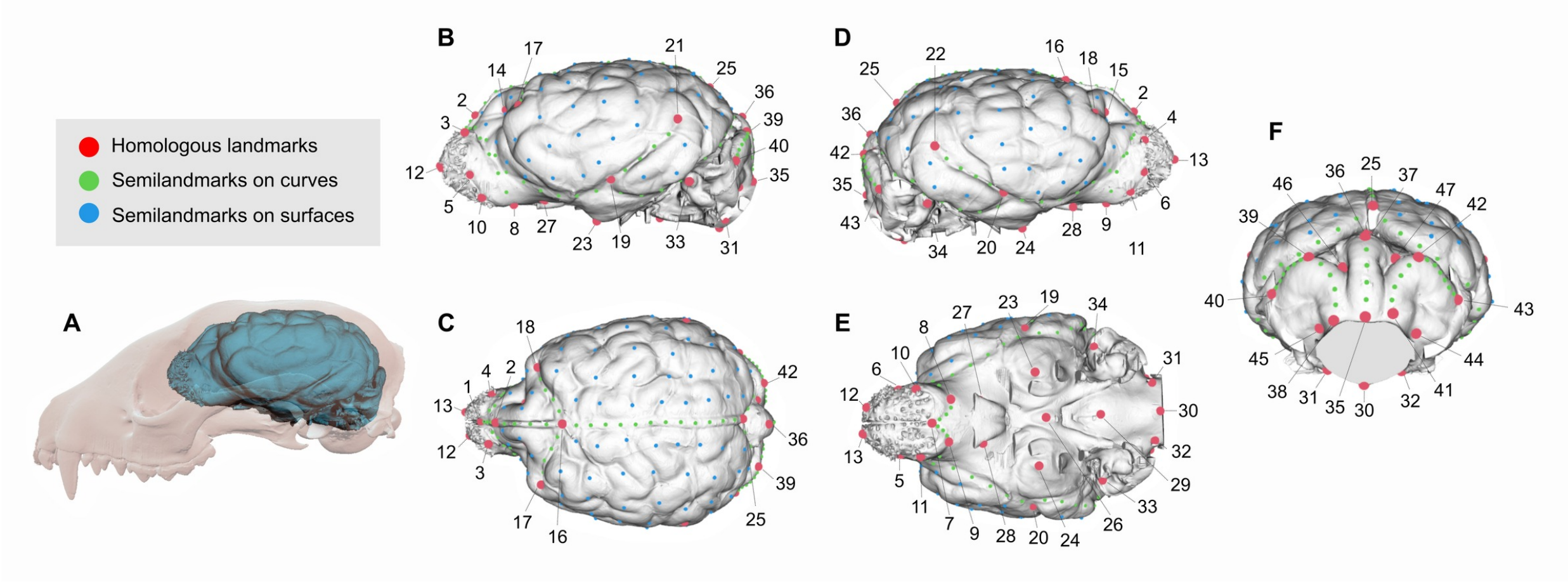

**Figure 2. Position of the landmarks captured on the endocast.** (**A**) Three-dimensional visualization of a common raccoon (*Procyon lotor*) cranium and endocast brain colored in blue. Landmarks and semilandmarks used in this study are visualized for the (**B**) left side; (**C**) dorsal view; (**D**) right side; (**E**) ventral view and (**F**) caudal view. Numbered red dots represent the anatomical landmarks, while green and blue dots represent the curves and the surface semilandmarks respectively.

**Table 1.** ***Table 1. Definition of the landmarks used to characterize the endocast shape.*** *The asterisked landmarks were defined following previous studies on mammalian endocast morphology (Ahrens, 2014; Bertrand et al., 2019).*

| Landmarks | Definition | Location |
|---|---|---|
| **1*** | Most antero-dorsal point of the olfactory bulbs, midline | Olfactory bulb |
| **2*** | Confluence of the circular and the sagittal fissure anterior, midline | Cerebrum |
| **3** | Dorsal terminus of the olfactory bulb, left | Olfactory bulb |
| **4** | Dorsal terminus of the olfactory bulb, right | Olfactory bulb |
| **5*** | Caudal constriction of the olfactory bulb at the dorsoventral midpoint, left | Olfactory bulb |
| **6** | Caudal constriction of the olfactory bulb at the dorsoventral midpoint, right | Olfactory bulb |
| **7*** | Most antero-ventral point of olfactory bulb, midline | Olfactory bulb |
| **8** | Most ventral point of contact between the olfactory bulb and the olfactory peduncle, left | Olfactory bulb |
| **9** | Most ventral point of contact between the olfactory bulb and the olfactory peduncle, right | Olfactory bulb |
| **10** | Ventral terminus of olfactory bulb, left | Olfactory bulb |
| **11** | Ventral terminus of olfactory bulb, right | Olfactory bulb |
| **12*** | Most anterior point of the olfactory bulb, left | Olfactory bulb |
| **13** | Most anterior point of the olfactory bulb, right | Olfactory bulb |
| **14*** | Anterior end of the presylvian sulcus, left | Cerebrum |
| **15** | Anterior end of the presylvian sulcus, right | Cerebrum |
| **16*** | Cruciate sulcus, midline | Cerebrum |
| **17** | Most lateral point of the cruciate sulcus, left | Cerebrum |
| **18** | Most lateral point of the cruciate sulcus, right | Cerebrum |
| **19** | Most ventral point of the pseudosylvian sulcus, left | Cerebrum |
| **20** | Most ventral point of the pseudosylvian sulcus, right | Cerebrum |
| **21*** | Most dorsal point of the pseudosylvian sulcus, left | Cerebrum |
| **22** | Most dorsal point of the pseudosylvian sulcus, right | Cerebrum |
| **23*** | Ventral terminus point of the piriform lobe, left | Cerebrum |
| **24** | Ventral terminus point of the piriform lobe, right | Cerebrum |
| **25*** | Caudal midline terminus between the left and right occipital lobes, midline | Cerebrum |
| **26*** | Hypophyseal fossa, midline | Cerebrum |

| Land-marks | Definition | Location |
|---|---|---|
| **27*** | Proximal-most differentiation of optic tract, left | Cerebrum |
| **28** | Proximal-most differentiation of optic tract, right | Cerebrum |
| **29** | Most ventral point of the pons, midline | Hindbrain |
| **30*** | Endobasion, midline | Hindbrain |
| **31** | Occipital emissary vein canal, left | Hindbrain |
| **32** | Occipital emissary vein canal, right | Hindbrain |
| **33** | Most lateral point of the hindbrain, left | Hindbrain |
| **34** | Most lateral point of the hindbrain, right | Hindbrain |
| **35*** | Ventral terminus of the vermis, midline | Hindbrain |
| **36*** | Dorsal terminus of the vermis, midline | Hindbrain |
| **37** | Most anterior point of contact between the cerebellum and the cerebrum, midline | Hindbrain |
| **38*** | Ventral terminus of the left cerebellar hemisphere | Hindbrain |
| **39*** | Dorsal terminus of the left cerebellar hemisphere | Hindbrain |
| **40*** | Lateral terminus of the left cerebellar hemisphere | Hindbrain |
| **41** | Ventral terminus of the right cerebellar hemisphere | Hindbrain |
| **42** | Dorsal terminus of the right cerebellar hemisphere | Hindbrain |
| **43** | Lateral terminus of the right cerebellar hemisphere | Hindbrain |
| **44** | Most ventro-lateral point of contact between the cerebellum and the medulla oblongata, right | Hindbrain |
| **45** | Most ventro-lateral point of contact between the cerebellum and the medulla oblongata, left | Hindbrain |
| **46** | Most dorsal point of the paravermis, left | Hindbrain |
| **47** | Most dorsal point of the paravermis, right | Hindbrain |

## Quantifying brain organization

To analyze brain composition at different organizational levels, we extracted two types of relative measurements from each specimen: (1) the volumes of the whole brain as well as the main anatomical regions, and (2) the surface areas of two functional cortical domains (i.e. the neocortex and the paleocortex). We focused on three major subdivisions of the endocast: the olfactory bulbs (OB), cerebrum (CR), and hindbrain (HB), which together represent key structural components of the mammalian brain (Kaas, 2017). The petrosal lobules represent another structure of interest for understanding adaptation to 3D environment (e.g. arboreal or aquatic locomotion); however, since many musteloids lack subarcuate fossae (i.e., cranial cavities that accommodate the petrosal lobules), they could not be consistently identified in all species and were thus not included in the present study.

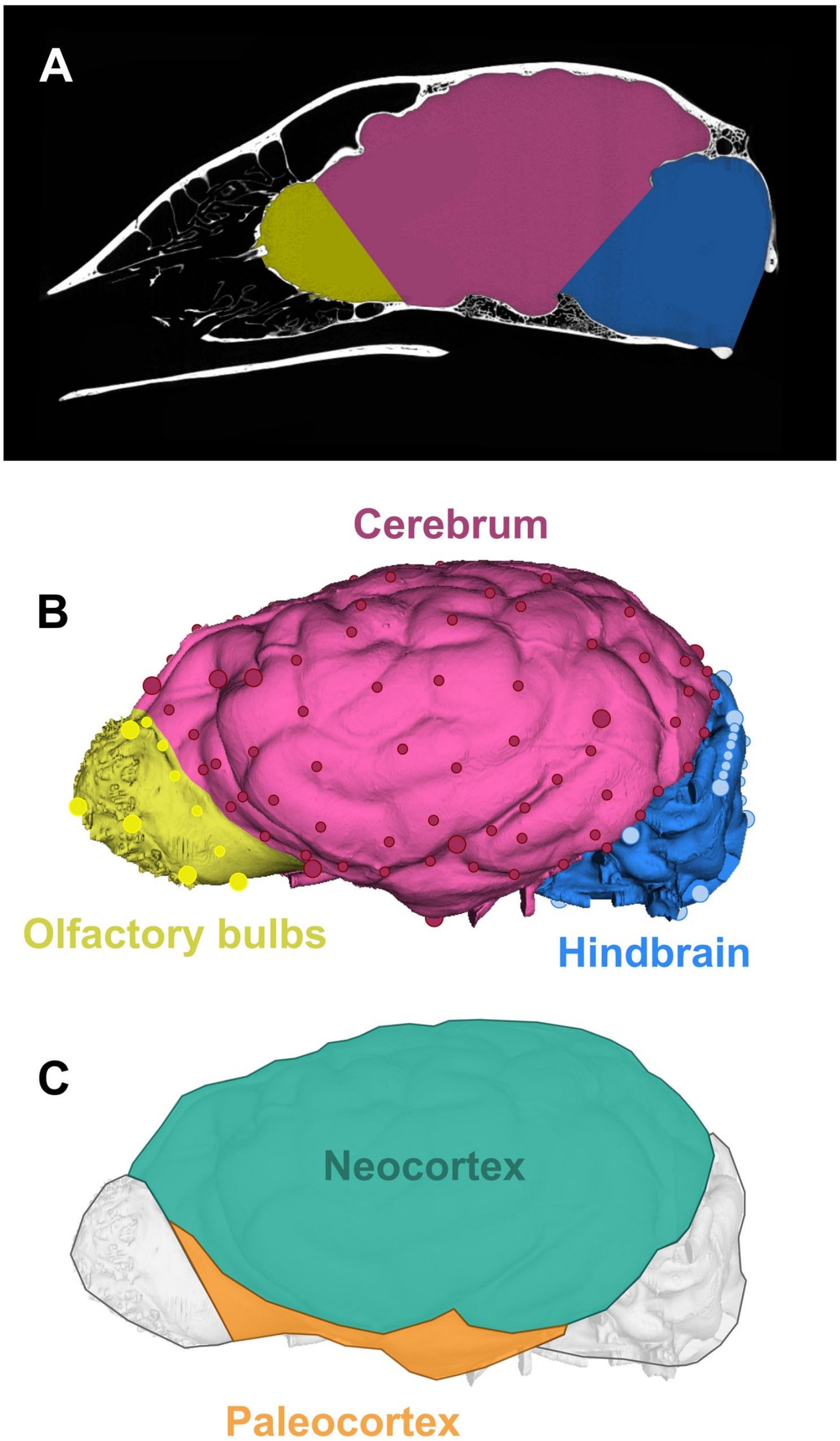

**Figure 3. Endocast regions analyzed in this study.** Endocast of a common raccoon (*Procyon lotor*) on **(A)** a sagittal section under AVIZO visualization; **(B)** the virtual 3D endocast extracted from this segmentation with the three main anatomical regions – the olfactory bulb (yellow), cerebrum (pink), and hindbrain (blue) – and their associated landmarks and semilandmarks and **(C)** the cortical regions corresponding to the neocortex (green) and the paleocortex (orange).

For each specimen, we first computed the centroid size of the whole endocast as well as that of each region (OB, CR and HB) as a measure of absolute regional size. In addition, the external endocast surface areas corresponding to the neocortex (associated with higher-order sensory and associative processing) and paleocortex (primarily linked to olfactory functions), were manually delineated on each specimen to quantify their surface areas on the 3D endocranial meshes. Neocortical boundaries were traced along the rhinal fissure, whereas the paleocortex was delineated by grouping the olfactory tracts, the piriform cortex, and the major cranial nerve impressions visible in this region of the endocast, following Bertrand et al. (2019). All surface area measurements are reported in **Supporting Information 1**. These volume and surface area values were then averaged per species and log-transformed when included in further analyses.

Finally, to assess the compositional structure of the brain, we treated the three regional volumes (OB, CR, HB) as components of a composition summing to 1 for each specimen, reflecting their proportional contribution to total brain size. Following the framework of compositional data analysis (Aitchison, 1982, 1986), these data were mapped into Euclidean space using an isometric log-ratio (ILR) transformation (Egozcue et al., 2003), which removes the constant-sum constraint inherent to proportions and enables classical multivariate testing within the Aitchison geometry. The ILR coordinates were then averaged by species in Aitchison space using the *aitchison_mean_by_group* procedure (Van den Boogaart & Tolosana-Delgado, 2013) and subsequently, back-transformed to retrieve the Aitchison mean composition. This transformation yields two orthogonal coordinates. The first axis ($z_1$) expresses the balance between OB and CR, capturing relative olfactory enlargement or reduction with respect to the rostral cerebral region. The second axis (z2) contrasts the geometric mean of OB and CR with HB, reflecting the relative allocation of tissue to anterior (OB+CR) versus posterior brain regions (HB). These ILR coordinates remove the constant-sum constraint and preserve the geometry of the simplex, thereby enabling standard multivariate analyses.

## Characterizing brain shape

All analyses were performed in R v4.5.0 (R Core Team, 2023) using RStudio v2023.06.1 (RStudio Team, 2020). First, we performed a Generalized Procrustes Analysis (GPA; Rohlf & Slice, 1990) using the *gpagen* function of the geomorph package v4.0.10 (Adams et al., 2021) for the entire brain shape as well as for each brain region (OB, CR and HB). For each dataset (entire brain, OB, CR and HB), overall size was quantified using centroid size extracted from the landmark configurations during GPA, and mean Procrustes coordinates and centroid sizes were then calculated per species.

We further investigated how these factors affect individual brain regions (OB, CR and HB) at a macroevolutionary scale. To explore patterns of shape variation in the whole endocast, we then performed phylogenetically informed Principal Component Analyses (PhyloPCA) on Procrustes-aligned landmark coordinates using the *gm.prcomp* function (Adams et al., 2021). Shape deformations along principal components (PCs) and regression trajectories were visualized using 3D warping of a reference mesh with the *warpRefMesh* function (Adams et al., 2021). The mean endocranial shape was used as the reference configuration for all deformation visualizations.

## Assessing the impact of taxonomic and ecological factors on brain organization and shape

In this study, we analyzed two complementary types of data as dependent variables: (1) relative size measures expressed as compositions (ILR-transformed regional volumes), surface areas (log-transformed), or proportions to assess how size, ecological categories, and taxonomic grouping impact brain organization; and (2) Procrustes-aligned shape data to evaluate how the same factors influence the morphology of the brain and its regions.

When multiple specimens were available for a given species, measurements were first computed for each individual and then averaged per species prior to phylogenetic analyses. Specifically, raw volumetric and surface area measurements were averaged across individuals, ILR coordinates were averaged in Aitchison space, and mean Procrustes coordinates and centroid sizes were calculated from the landmark configurations after Generalized Procrustes Analysis.

We also used two complementary size variables: endocast centroid size (referred as Csize) and body mass, as they capture distinct and biologically meaningful aspects of scaling. Centroid size provides a direct geometric proxy for brain size, whereas body mass reflects overall organismal size and the allometric constraints associated with it. Including both measures improves interpretability by allowing us to disentangle brain-specific size variation from variation driven by general body scaling. Both variables were log-transformed prior to the analyses. Similarly, analyses were performed using both Family and Subfamily as taxonomic variables to distinguish taxonomic effects across hierarchical levels.

All analyses followed a common analytical design that explicitly accounted for phylogenetic non-independences. The carnivoran phylogeny (modified from Hassanin et al. (2021)) shown in **Supporting Information 2** was pruned to match our species sample using the *ape* package (Paradis et al., 2004). For univariate data (e.g., brain proportions and surface areas), we used Blomberg's K (Blomberg et al., 2003) and Pagel's λ (Pagel, 1999). Blomberg's K quantifies the degree to which trait similarity among species indicates their phylogenetic relatedness, with K = 1 indicating a perfect Brownian motion pattern, K < 1 suggesting higher divergence than expected under Brownian motion, and K > 1 indicating stronger trait conservatism. Meanwhile, Pagel's λ measures the phylogenetic dependence of trait variation, with values ranging from 0 (i.e., phylogenetically independent) to 1 (i.e., Brownian motion). Significance was assessed via 1000 random permutations. For multivariate data (e.g., shape data), we quantified the phylogenetic signal using $K_{mult}$ (Adams & Felice, 2014) as implemented in *geomorph*. $K_{mult}$ assesses the fit of observed shape variation to a Brownian motion model of evolution via 1000 permutations.

The analytical framework consisted of three main steps: (i) a backward stepwise elimination procedure to identify the best-supported model, (ii) phylogenetic regression analyses (i.e., Phylogenetic Generalized Least Squares, referred as PGLS) performed on each factor retained in the best model, including pairwise comparisons, and (iii) an allometric trajectory analysis for the factors that showed significant effects. All models were run with 1000 residual randomization permutations to estimate significant levels robustly (Adams & Collyer, 2018). Our framework is detailed as follows:

(i) To assess how size, ecological variables, and taxonomy influence each univariate response variable (Y), we implemented a two-step phylogenetic model-selection procedure based on PGLS similar to the method provided by Gálvez-López (2021). Because both size and taxonomy can be defined in alternative ways, all analyses were repeated using two size metrics (Csize and body mass) and two taxonomic variables (Family and Subfamily). For each response variable Y and for each combination of size metric and taxonomic variable, we first fitted a full additive PGLS model including Diet, Locomotion, the chosen taxonomic factor (Family or Subfamily), and the selected size metric (Csize or body mass). Analyses were implemented using the *procD.pgls* function (Adams et al., 2021). We then applied a backward stepwise elimination to identify the best-supported set of main effects. At each iteration, non-significant predictors were ranked according to their Type III p-values and partial $\eta^2$ effect sizes (partial $\eta^2 < 0.01$), and the least informative term was removed. At each step, the removable term with the smallest $\eta^2$ was dropped, the model was refitted, and the procedure repeated. Backward elimination continued until no term met the removal criteria, yielding a final reduced model containing only statistically supported or non-negligible effects. Because including all possible interactions in the stepwise procedure would greatly increase model complexity and reduce power, interactions were tested separately. Although interactions can sometimes be statistically significant even when their corresponding main effects are not, we first identified the best additive model to limit the number of candidate interactions. Each interaction was then added individually to the additive model and evaluated using PGLS, regardless of the statistical significance of its main effects in the additive model. Interactions were retained only when they were statistically significant and showed non-negligible effect size (partial $\eta^2 \geq 0.01$); if several met these criteria, the interaction with the largest effect was selected.

To assess whether our conclusions depended on the backward stepwise procedure, we performed two complementary robustness checks. First, we evaluated the full additive model including size, taxonomic group, diet, and locomotion for all response variables, allowing us to report effect sizes and significance for all predictors. Second, we ranked the same additive candidate model set (i.e., all additive combinations of the four response variables in the full model) using AICc as an alternative model-selection criterion. Because AICc-based GLS ranking is not directly appropriate for multivariate Procrustes shape responses, robustness of shape analyses was instead assessed by exhaustive procD.pgls comparisons of goodness-of-fit across the same additive candidate sets (i.e., explained variance, residual sum-of-squares). Backward stepwise models consistently fell within the set of top-ranked candidate models ($\Delta$AICc $< 2$) or were aligned with the best model candidate set (similar goodness-of-fit), meaning that all approaches yielded similar broad conclusions. The detailed results of these additional analyses are presented in **Supporting Information 4**.

(ii) After identifying which predictors were statistically supported in the stepwise model-selection procedure, we carried out targeted phylogenetic regressions (PGLS) to examine their effects more explicitly. When PGLS analyses revealed a significant predictor or interaction effect (i.e., size*predictor), we conducted pairwise comparisons to identify which groups differed significantly. For each retained dependent variable - predictor combination, we estimated marginal means under the corresponding PGLS model with 1000 residual randomization permutations and performed post-hoc Tukey-adjusted pairwise contrasts to compare levels of the categorical factor. To ensure statistical reliability, only factor levels with a minimum of three species per group ($n \geq 3$) were preserved in pairwise tests.

(iii) Finally, to evaluate patterns of allometric scaling, we examined how our size metrics (centroid size and body mass) interacted with three types of brain descriptors: compositional variables (ILR coordinates), proportional measures (relative sizes of OB, CR and HB), and cortical surface areas (neocortex and paleocortex). These relationships were analyzed using multivariate phylogenetic GLS models (*procD.pgls*), testing the size × predictor interaction separately for each size metric and each descriptor. When this interaction was significant, we then compared allometric slopes among ecological or taxonomic groups to identify which groups deviated from the global allometric trajectory. To formally quantify these differences in slopes and intercepts, we fitted Bayesian hierarchical linear models using the *rstanarm* package (Goodrich et al., 2022), running on the Stan probabilistic platform (Carpenter et al., 2017). This approach is well suited to our dataset, which contains small and uneven sample sizes and phylogenetically structured observations. Prior to model fitting, the size variable (Csize or body mass) was mean-centred to improve MCMC sampling efficiency and stabilize slope estimation. Species-level random intercepts and random slopes were modelled using a non-correlated parametrization, following best practices in multilevel Bayesian modelling (Gelman & Hill, 2007). Slightly informative priors were used, and sampling was performed with stricter NUTS settings to reduce divergent transitions (delta = 0.999). For each analysis, we first estimated a global allometric slope across all species, which served as a reference. Differences between each group's slope and the global slope were quantified using posterior means, 95% credible intervals, and Bayesian tail probabilities. For shape data, we applied the same Bayesian framework to a reduced residual shape space built from PCs explaining >80% of the total shape variance (first five PCs). For each PC, we fitted a hierarchical Bayesian regression of PC score on mean-centered size using the same model structure described above. From the posterior samples of all five models, we reconstructed multivariate allometric vectors (one slope per PC) for the global model and for each ecological group. This allowed us to quantify group differences not only in the magnitude of residual-shape allometry, but also in the orientation of allometric trajectories in the multivariate PC space, using posterior distributions of vector lengths and angular deviations. This framework provides a robust evaluation of group-level deviations in allometric scaling while fully propagating uncertainty and accounting for hierarchical and phylogenetic structure.

# RESULTS

All results reported below are based on species-level analyses (n = 41), using species means derived from individual measurements for volumetric and surface variables, ILR coordinates, and Procrustes-aligned landmark configurations.

## Description of brain organization and shape in musteloids

Our results reveal substantial variation in brain composition across species in our dataset (**Figure 4, Table 2**). When examining the relative proportions of the three major anatomical components (OB, CR, and HB), two groups stand out clearly: the Lutrinae (a subfamily within Mustelidae) and the Mephitidae. Lutrinae shows the largest CR proportion (74.3%) and the smallest proportions for both OB (8.6%) and HB (17.1%). In contrast, Mephitidae displays notably larger OB (14.5%) and the smallest proportion of CR (66.3%) among all families and subfamilies. Additionally, arboreal species exhibit the highest relative proportion of hindbrain tissue (19.8%) compared to all other groups. These patterns are descriptive, and the contribution of ecological variables is formally assessed in the phylogenetic models presented below..

**Table 2.** *Table 2. Mean relative contributions of major brain regions (OB, CR, HB) across ecological and taxonomic categories. Values represent descriptive summaries and do not account for phylogeny or size effects.*

| Predictor | Group | OB | CR | HB |
|---|---|---|---|---|
| **Family** | Ailuridae | 9.2% | 72.2% | 18.6% |
| | Mephitidae | 14.5% | 66.3% | 19.2% |
| | Mustelidae | 10.9% | 70.8% | 18.3% |
| | Procyonidae | 10.8% | 70.5% | 18.7% |
| **Subfamily** | Ailuridae | 9.19% | 72.2% | 18.6% |
| | Lutrinae | 8.6% | 74.3% | 17.1% |
| | Guloninae + Helictidinae | 11.6% | 69.9% | 18.4% |
| | Ictonychinae | 11.8% | 68.9% | 19.3% |
| | Mustelinae | 12.3% | 68.2% | 19.5% |
| | Mephitidae | 14.5% | 66.3% | 19.2% |
| | Procyonidae | 10.8% | 70.5% | 18.7% |
| | Badgers | 11.4% | 70.6% | 18.0% |
| **Diet** | Herbivorous | 10.2% | 70.6% | 19.2% |
| | Omnivorous | 11.9% | 69.5% | 18.5% |
| | Piscivorous | 9% | 73.5% | 17.5% |
| | Carnivorous | 12.0% | 68.8% | 19.2% |
| **Locomotion** | Semi-aquatic | 8.9% | 73.7% | 17.4% |
| | Arboreal | 11.2% | 69.0% | 19.8% |
| | Semi-arboreal | 11.0% | 70.6% | 18.3% |
| | Semi-fossorial | 12.2% | 69.6% | 18.2% |
| | All-terrain | 12.5% | 68.0% | 19.5% |
| | Scansorial | 11.2% | 70.3% | 18.5% |

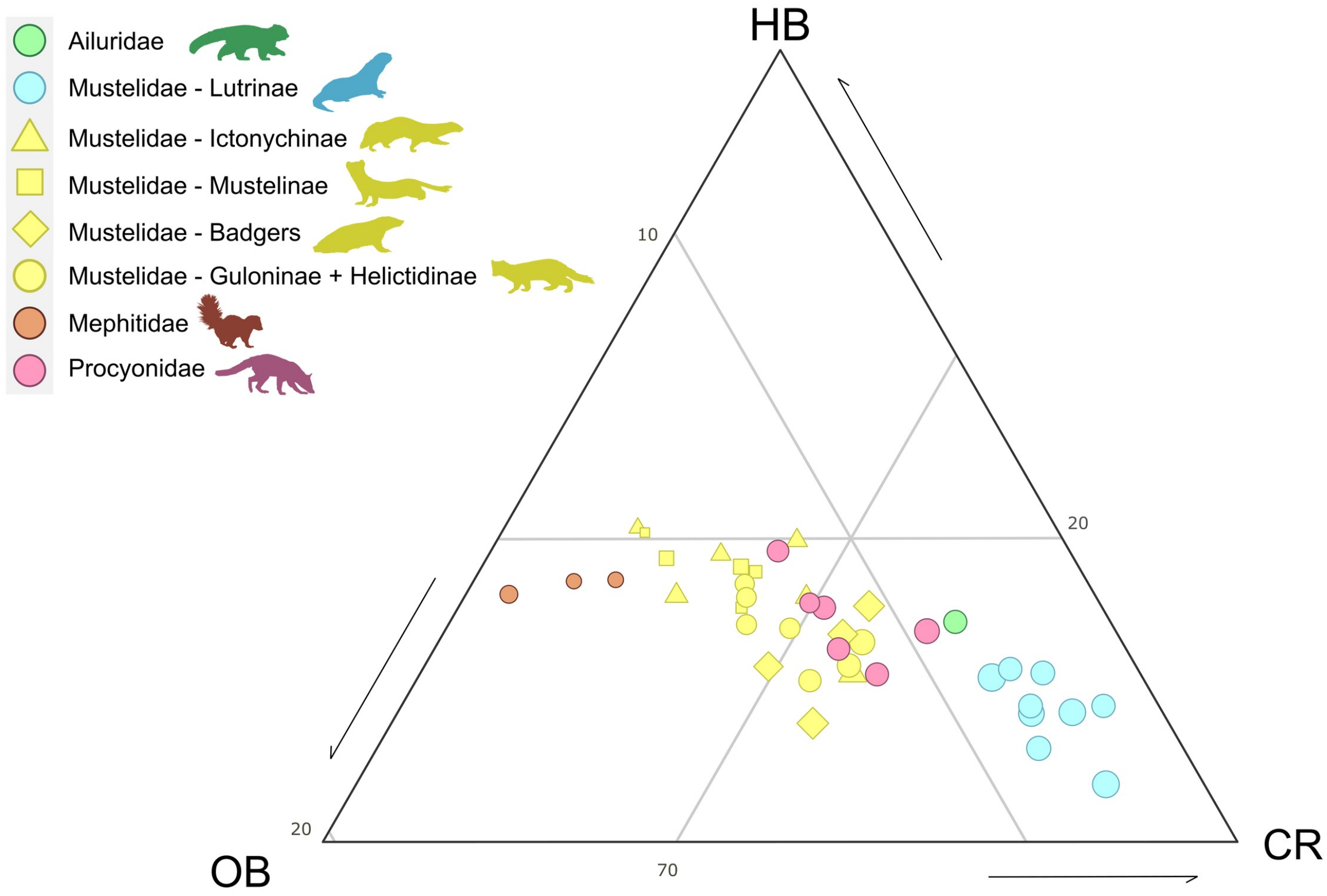

**Figure 4. Ternary diagram of relative brain region proportions across taxonomic groups.** Point size reflects log-transformed endocranial size. The arrows indicate the direction of increasing contribution for each component of the ternary space.

The PhyloPCA revealed that the first two principal components (PCs) accounted for 43.3% of the overall shape variance (**Figure 5**). The first axis, accounting for 14.7% of the overall variance, tended to separate species characterized by a rounder, medio-laterally wider and antero-caudally compressed endocast with ventrally oriented olfactory bulbs (positive part of PC1) from species characterized by a relatively narrower endocast with rostrally oriented olfactory bulbs and a dorso-ventrally flatter cerebrum (negative part of the PC1). Nevertheless, PC1 does not differentiate taxonomic or ecological groups according to their endocast morphology. As with the first axis, the distribution defined by the scatter plot along PC2 (10.2% of the overall shape variance) does not reveal a clear separation between the different taxonomic groups. The morphological variation along this axis reflects a gradient from species with a rounder cerebrum on both the antero-caudal and dorsoventral axes and a small olfactory bulb (positive side of PC2) to those with a more elongated endocast along the antero-caudal axis, proportionally larger olfactory bulbs and a more prominent piriform lobe (negative side of PC2).

## Phylogenetic signal

All traits describing brain composition showed significant phylogenetic signal according to both Blomberg's K and Pagel's $\lambda$, indicating that closely related species tend to resemble one another in these traits. (**Table 3**). The only exception was the second ILR axis which exhibited a weaker and mixed pattern, with a significant K ($K = 0.378$, $p$-value $= 0.035$) but a non-significant $\lambda$ ($\lambda = 0.641$, $p$-value $= 0.109$). This suggests that the contrasted allocation between anterior brain regions (OB + CR) and the posterior region (HB) retains only a modest phylogenetic structuring: closely related species resemble each other slightly more than expected by chance, yet the overall distribution of this anatomical contrast does not closely follow the phylogenetic topology.

Shape data similarly exhibited significant phylogenetic signal for the entire brain (K = 0.59, p = 0.001), as well as for the shapes of the OB (K = 0.44, p = 0.001), CR (K = 0.54, *p*-value = 0.001), and HB (K = 0.54, *p*-value = 0.001).

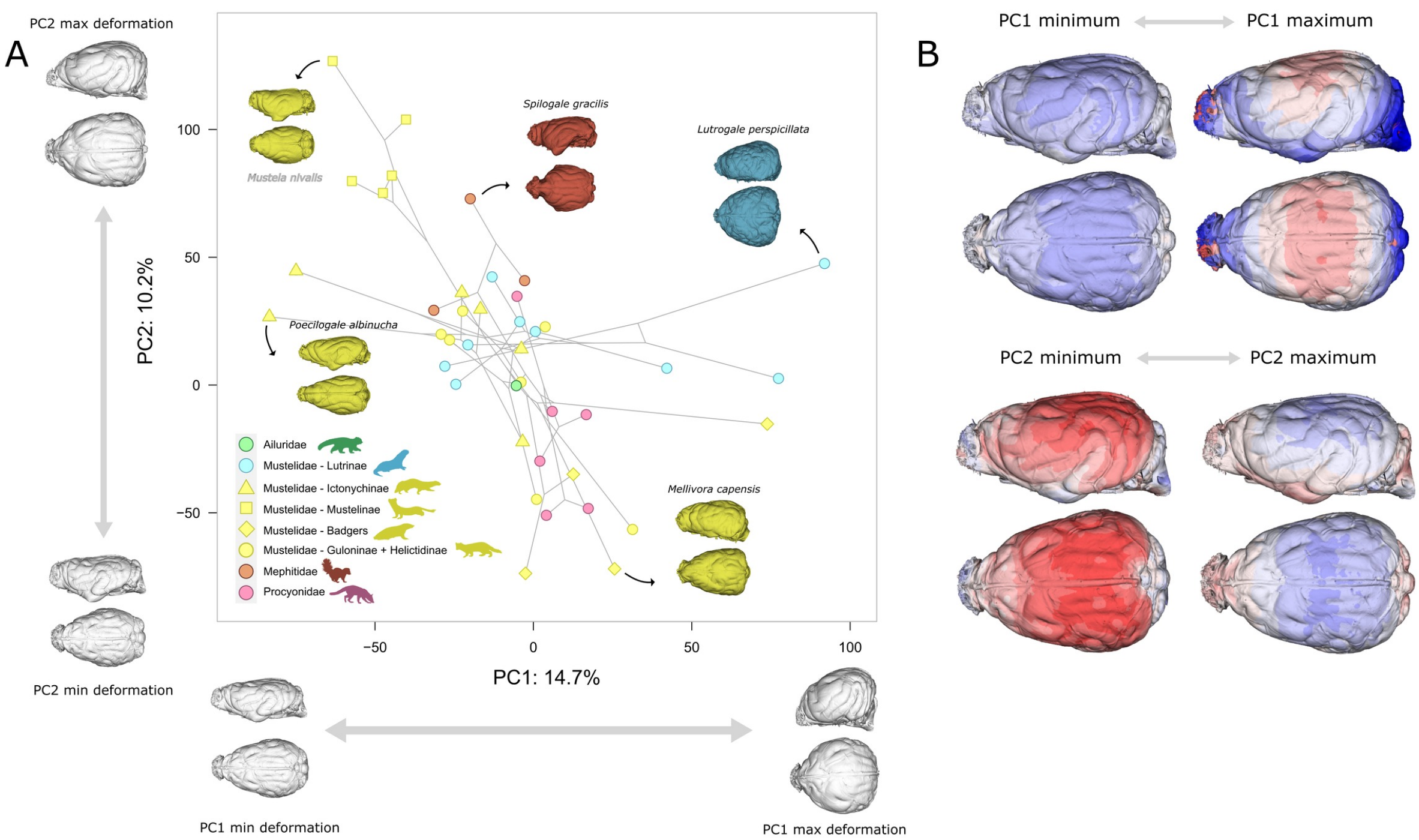

**Figure 5. Morphological variation in the musteloid endocast.** (**A**) PhyloPCA on the endocast shape and (**B**) shape variation along the first and second PCs obtained from warped meshes. Red areas indicate relative expansions from the mean conformation whereas blue areas represent relative contractions. A few outlier species were labeled on the plots for clarity.

**Table 3.** ***Table 3. Phylogenetic signal estimated for all traits describing brain composition using Blomberg's K and Pagel's λ.*** *Significance of K was evaluated through randomization tests, whereas λ significance was assessed via likelihood comparison with a λ = 0 model.*

| Trait | Description | K | K p-value | λ | λ p-value |
|---|---|---|---|---|---|
| ILR z1 | Olfactory changed with respect to the rostral cerebral region | 1.41 | <0.001*** | 0.98 | <0.001*** |
| ILR z2 | Balance between anterior versus posterior brain regions | 0.38 | 0.035* | 0.64 | 0.110 |
| OB | Proportion of the olfactory bulbs | 1.40 | <0.001*** | 0.98 | <0.001*** |
| CR | Proportion of the cerebrum | 1.55 | <0.001*** | 1 | <0.001*** |
| HB | Proportion of the hindbrain | 0.79 | <0.001*** | 1 | <0.001*** |
| Paleocortex | Surface area of the paleocortex | 0.59 | 0.002** | 0.88 | <0.001*** |
| Neocortex | Surface area of the neocortex | 0.87 | <0.001*** | 0.96 | <0.001*** |

## Ecology and taxonomy as determinants of brain structure and shape

We found a heterogeneous pattern in the set of the best-supported models identified by our backward stepwise PGLS procedure (**Table 4**). Importantly, the same best model was selected for each trait regardless of the size metric used (Csize or body mass). Across traits, three predictors emerged as key determinants of brain composition: the size metric itself, diet, and taxonomic grouping at the subfamily level.

The second ILR axis (reflecting the balance between anterior and posterior brain regions) was influenced by both diet and size, accounting for 31–35% of the total variance. For the proportion of HB, as well as for the neocortex and paleocortex surface areas, size was the only retained predictor, and this remained true regardless of whether Csize or body mass was used. These relationships were particularly strong for the neocortex, with the best models explaining up to 98% of the variance (84% when considering body mass), followed by the paleocortex (87% -83%), and the hindbrain (35-32%). By contrast, the first ILR axis (capturing olfactory shifts relative to the rostral cerebrum) and the proportions of OB and CR were primarily shaped by taxonomy, particularly at the subfamily level, with these models explaining between 43%-50% of total variance. When analyses were repeated at the broader family level, the first ILR axis and OB were instead best explained by diet, whereas CR became exclusively size dependent. In all three cases, the explained variance dropped substantially compared to the subfamily-based models, reaching only 17% to 30% depending on the analyses.

Across all shape analyses, the best-supported models included size as the only significant predictor, regardless of whether size was represented by endocast centroid size or body mass. This pattern was consistent for both whole-brain shape and the shape of individual brain structures ($p$-value <0.03 for all models on shape data). However, size accounted for only a small proportion of the observed shape variation, explaining approximately 5–8% of the total variance. Detailed model outputs are provided in **Supporting Information 5**.

**Table 4.** ***Table 4. Best-supported PGLS models for each brain composition trait.*** *(A) Results based on subfamilies as the taxonomic grouping. (B) Results based on families as the taxonomic grouping; only models that differ from the subfamily-based analyses are reported.*

**(A) Analyses using subfamilies as the taxonomic grouping**

| Structure | Formula (best model) | R² (Csize) | p-value (Csize) | R² (Body mass) | p-value (Body mass) |
|---|---|---|---|---|---|
| ILR z1 | Y ~ Subamily | 0.44 | 0.01** | 0.44 | 0.01** |
| ILR z2 | Y ~ Diet + size | 0.31 | Diet=0.015*/ Size=0.006** | 0.35 | Diet=0.024*/ Size=0.001*** |
| OB proportion | Y ~ Subfamily | 0.43 | 0.012* | 0.43 | 0.012* |
| CR proportion | Y ~ Subfamily | 0.50 | 0.002** | 0.50 | 0.002** |
| HB proportion | Y ~ Size | 0.35 | 0.002** | 0.32 | 0.001*** |
| Neocortex | Y ~ Size | 0.98 | 0.001** | 0.84 | 0.001** |
| Paleocortex | Y ~ Size | 0.87 | 0.001*** | 0.83 | 0.001*** |
| Brain shape | Y ~ Size | 0.07 | 0.007** | 0.07 | 0.015* |
| OB shape | Y ~ Size | 0.07 | 0.002* | 0.08 | 0.001*** |
| CR shape | Y ~ Size | 0.05 | 0.03* | 0.06 | 0.03* |
| HB shape | Y ~ Size | 0.05 | 0.03* | 0.06 | 0.03* |

**(B) Analyses using families as the taxonomic grouping**

| Structure | Formula (best model) | R² (Csize) | p-value (Csize) | R² (Body mass) | p-value (Body mass) |
|---|---|---|---|---|---|
| ILR z1 | Y ~ Diet | 0.22 | 0.024* | 0.22 | 0.024* |
| OB proportion | Y ~ Diet | 0.23 | 0.018* | 0.23 | 0.018* |
| CR proportion | Y ~ Size | 0.30 | 0.001*** | 0.17 | 0.008** |

Agreement between model-selection approaches varied across response-variable types but remained broadly consistent with the main conclusions (see **Supporting Information 4**). For multivariate brain shape variables and brain volumes, both approaches showed full agreement, consistently identifying size as the primary predictor. For proportional metrics, agreement was partial, as backward stepwise models sometimes retained taxonomic effects whereas AICc ranking favored simpler models dominated by size or diet. Similarly, for ILR-transformed components, agreement was moderate, with back-ward stepwise models often including diet in addition to size while AICc-supported models tended to emphasize size alone or in combination with other predictors. Importantly, in all univariate cases the backward stepwise models fell within the set of candidate models with ΔAICc < 2, indicating that both approaches yielded statistically equivalent model structures and supporting the robustness of the main results.

## Pairwise differences between groups

All detailed pairwise PGLS outputs, including full ANOVA tables, estimated marginal means, and complete Tukey contrasts for each analysis, are provided in **Supporting Information 6.**

Pairwise post-hoc PGLS comparisons performed on the predictors retained in the best models revealed several coherent patterns. For the first ILR 1 (ILR z1), significant differences were found between Lutrinae and all other subfamilies (t = –2.75 to –5.78; *p*-value < 0.002), except for Procyonidae (*p*-value = 0.119). These contrasts indicate that Lutrinae displays markedly lower ILR z1 values, reflecting OB proportion compared with CR. This pattern was confirmed in the pairwise analyses conducted directly on OB, where Lutrinae again differed significantly from all other groups (t = –3.68 to –4.78; *p*-value < 0.013), except Procyonidae (*p*-value = 0.375). Interestingly, an additional contrast emerged between Mephitidae and Procyonidae (t = 3.246; *p*-value = 0.039), consistent with proportionally larger OB proportion in Mephitidae relative to Procyonidae. A similar pattern to OB proportion was recovered for CR, with Lutrinae differing significantly from all other subfamilies (t = 3.23 to 4.73; *p*-value < 0.04), except again for Procyonidae (*p*-value = 0.16). In this case, however, the direction of the effect was reversed: Lutrinae exhibited proportionally larger CR compared with the other clades.

For the second ILR axis (ILR z2), significant differences were detected between omnivores and piscivores across both size metrics (Csize: t = 2.556, *p*-value = 0.039; body mass: t = 2.603, *p*-value = 0.035). Because ILR z2 reflects the balance between anterior (OB + CR) and posterior (HB) brain regions, these positive contrasts indicate that piscivores allocate relatively more tissue to the hindbrain compared with the anterior regions than do omnivores.

When analyses were repeated at the family level rather than at the subfamily level, a consistent pattern was found for both ILR z1 and OB. ILR z1 differed significantly between piscivores and both omnivores and carnivores (t = –3.15 and –3.44, respectively; *p*-value < 0.009), indicating again lower ILR z1 values in piscivores. Pairwise comparisons on OB showed the same contrasts, with piscivores differing significantly from omnivores and carnivores (t = –2.98 and –3.21, respectively; *p*-value < 0.014). Together, these results consistently show that piscivores, including both otters and other musteloids, possess proportionally smaller OB relative to the rostral cerebrum (CR) compared with other dietary groups.

## Allometric scaling

All our brain composition variables show a strong correlation with the size metrics for both Csize and body mass (**Figure 6**), except for ILR axis 2 (Csize: *p*-value = 0.752; body mass: *p*-value = 0.432). This indicates that the balance between the anterior (OB + CR) and posterior (HB) brain region changes independently from both endocranial size or body mass within musteloids. For the three main anatomical regions, our results highlight that CR proportion increases significantly with size (Csize : $R^2$ = 0.66, body mass: $R^2$ = 0.52; *p*-value <0.001), while both OB and HB proportions decrease (OB, Csize : $R^2$ = 0.65; body mass : $R^2$ = 0.58; *p*-value <0.001 ; HB, Csize : $R^2$ = 0.50; body mass : $R^2$ = 0.36; *p*-value <0.001). Neocortex and paleocortex surface areas showed similar scaling trends, both exhibiting strong positive correlations with size (Csize: $R^2$ = 0.99, body mass: $R^2$ = 0.90; *p*-value <0.001).

For all structures, we consistently found a stronger correlation when brain size (Csize) was used as the size metric rather than body mass, which would be expected given their direct anatomical relationship with the brain. However, an exception was observed for paleocortex surface area, which showed a slightly higher $R^2$ when body mass was considered as the scaling variable (Csize: $R^2 = 0.90$; body mass: $R^2 = 0.91$; *p*-value <0.001).

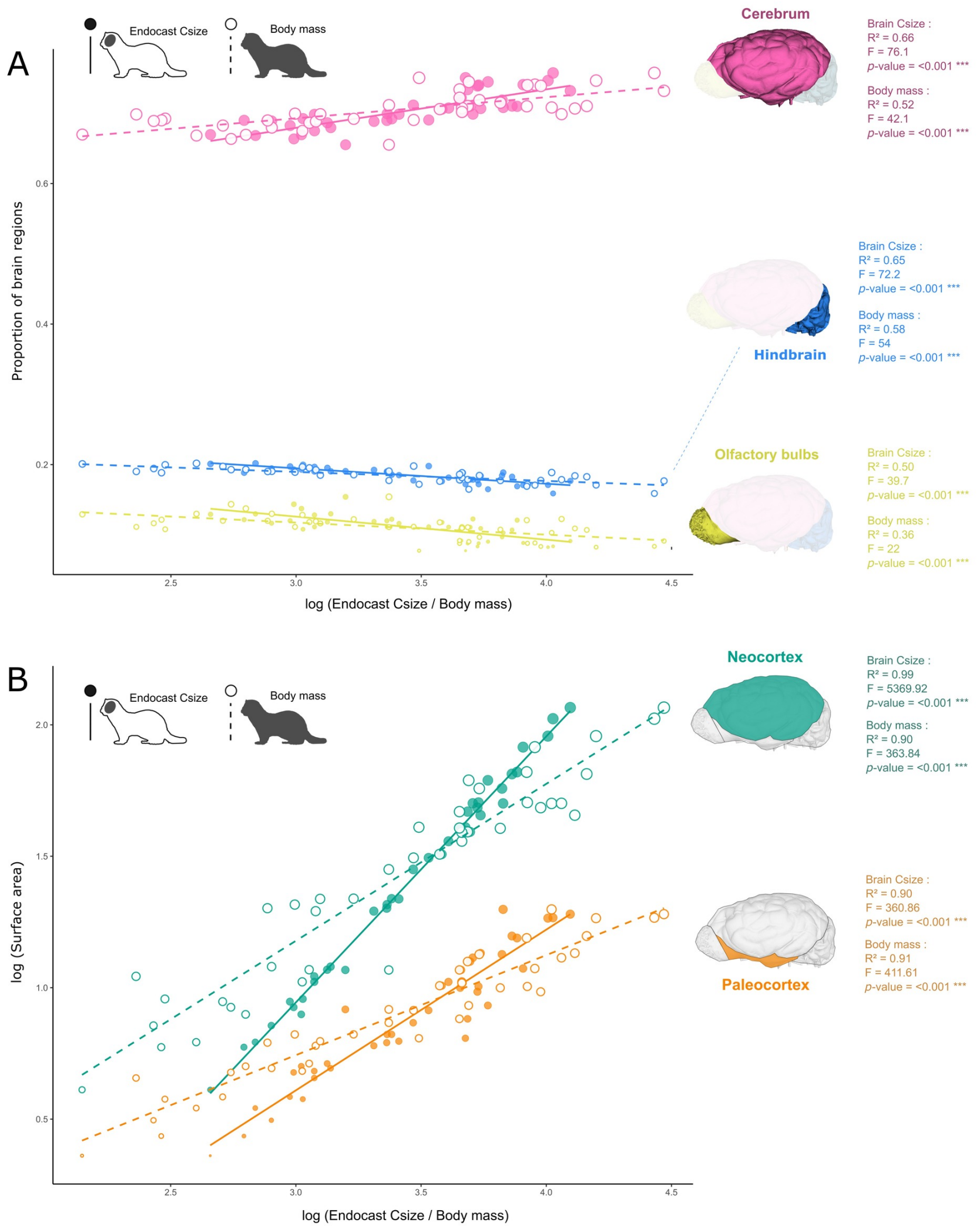

**Figure 6. Allometric relationships between brain regions and size in musteloids.** (**A**) Relationships between each brain region proportion (OB: yellow, CR: pink, HB: blue) and (**B**) log-transformed surface areas of the neocortex (green) and the paleocortex (orange) against the log-transformed metric size (endocastCsize and body mass) across musteloid species. Point size reflects regional percentage. Solid regression lines with filled markers indicate the relationships estimated using endocast centroid size, whereas dashed lines with unfilled markers represent relationships estimated using body mass.

PGLS models including only size as a covariate revealed a consistent but modest allometric signal in brain shape variation. Whole-brain shape showed a significant association with size, whether measured as endocranial centroid size ($R^2 = 0.07$, *p*-value = 0.007) or body mass ($R^2 = 0.07$, *p*-value = 0.015). Comparable effects were detected for OB, CR and HB shape, regardless of which size metric was used (*p*-value <0.03 for all brain regions). Nevertheless, in all cases size accounted for only a small proportion of shape variation ($R^2$ = 0.05–0.08), indicating weak allometric constraints on neuroanatomical form in Carnivora.

Across all analyses incorporating interaction terms between size metrics and ecological and taxonomic predictors, significant interactions were only recovered for paleocortex surface area and both endocast and regional shape (**Supporting Information 7**). Indeed, paleocortical surface exhibited taxon-specific allometries: its scaling relationship with size differed significantly among subfamilies whether we used Csize ($R^2 = 0.11$, *p*-value = 0.002) or body mass ($R^2 = 0.18$, *p*-value = 0.001). When size was expressed as body mass, additional significant interactions appeared with ecological factors, including both diet ($R^2 = 0.05$, *p*-value = 0.05) and locomotion ($R^2 = 0.09$, *p*-value = 0.02) with the latter showing a comparatively stronger effect. These results indicate that the paleocortex does not scale uniformly across ecological groups: its allometric trajectory varies with both taxonomic lineage and ecological niche. A similar pattern was detected for the neocortex surface as PGLS model revealed significant differences in neocortical surface scaling among subfamilies when size was expressed as body mass ($R^2 = 0.23$, *p*-value = 0.047).

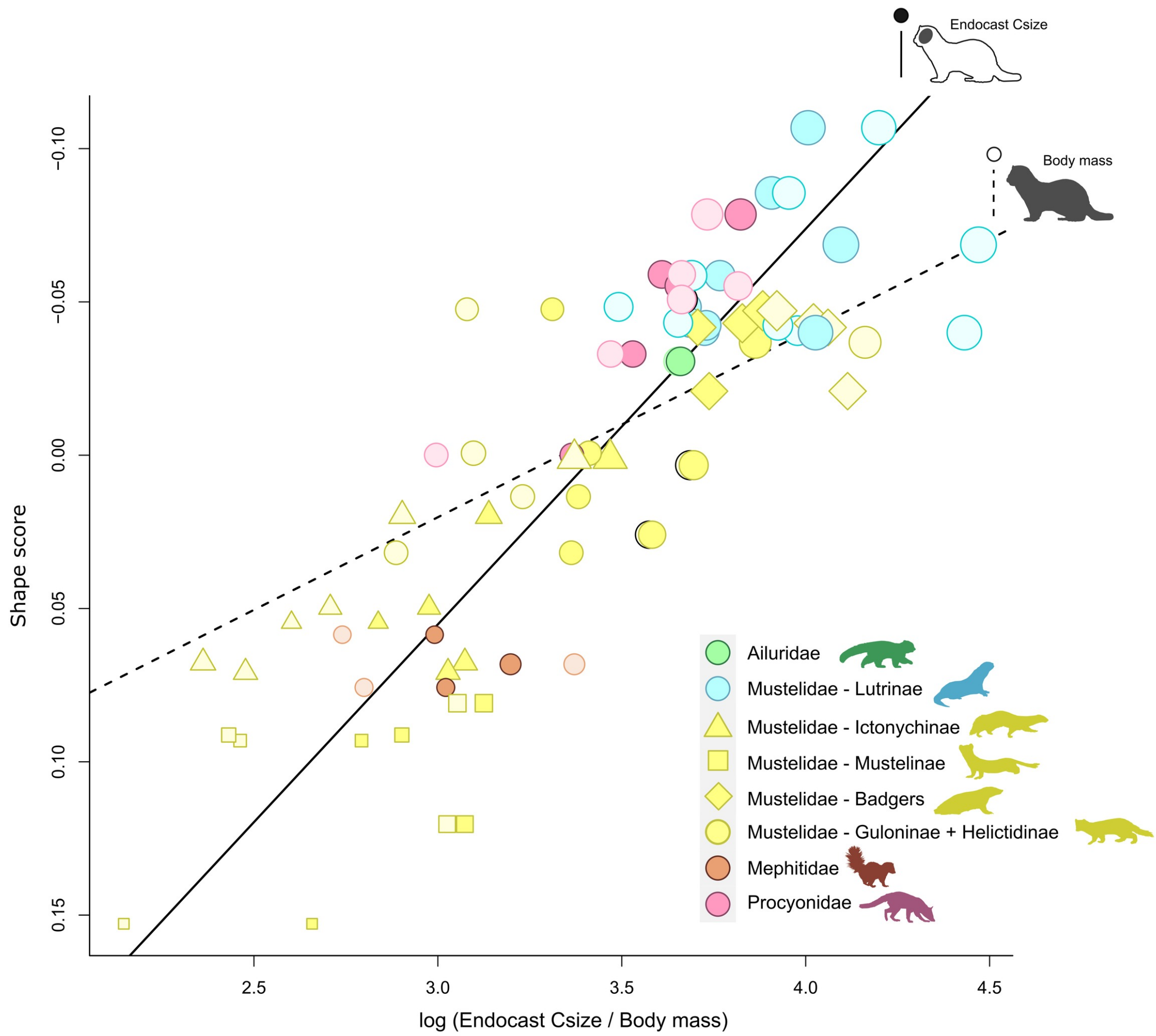

**Figure 7. Shape regression scores against the endocast centroid size and body mass in musteloids.** Point size is scaled to the log-transformed size metrics (endocast Csize and body mass). Solid regression lines with filled markers indicate the relationships estimated using endocast centroid size, whereas dashed lines with lighter-coloured markers represent relationships estimated using body mass.

The Bayesian hierarchical models provided further insight into variation in paleocortex surface area across Musteloidea. When accounting for brain size (Csize) several subfamilies showed allometric slopes that differed significantly from the global slope. This is particularly evident in Lutrinae (*p*-value < 0.001), Guloninae + Helictidinae (*p*-value = 0.018), and Mephitidae (*p*-value = 0.44; **Figure 8A**). We also detect differences in the orientation of the allometric relationships for CR proportion specifically in Lutrinae (*p*-value = 0.016) and Mephitidae (*p*-value = 0.013; **Figure 8B**). However, results for Mephitidae should be interpreted with caution: this group includes only three species in our dataset, and the observed divergence may therefore reflect sampling limitations rather than a strong evolutionary signal. Complete model outputs are provided in **Supporting Information 8**.

Regarding shape, significant between-group differences emerged only when considering taxonomic grouping. In the interaction models including subfamily, we detected significant allometric differences among subfamilies for whole-brain shape ($R^2 = 0.50$, *p*-value = 0.03), as well as for CR (CR; $R^2 = 0.48$, *p*-value = 0.034) and HB shape ($R^2 = 0.48$, *p*-value = 0.034) when Csize was used as the covariate. However, no such interaction was detected for OB shape nor in any models using body mass as the size predictor, suggesting that the detected shape differences relate more directly to brain size rather than to overall body scaling. Bayesian hierarchical models did not reveal any statistically supported differences in allometric slopes among subfamilies, indicating that the direction and magnitude of shape allometry appear largely conserved across the group. The full set of PGLS outputs, covering all interaction models between size and each predictor, is available in **Supporting Information 7.**

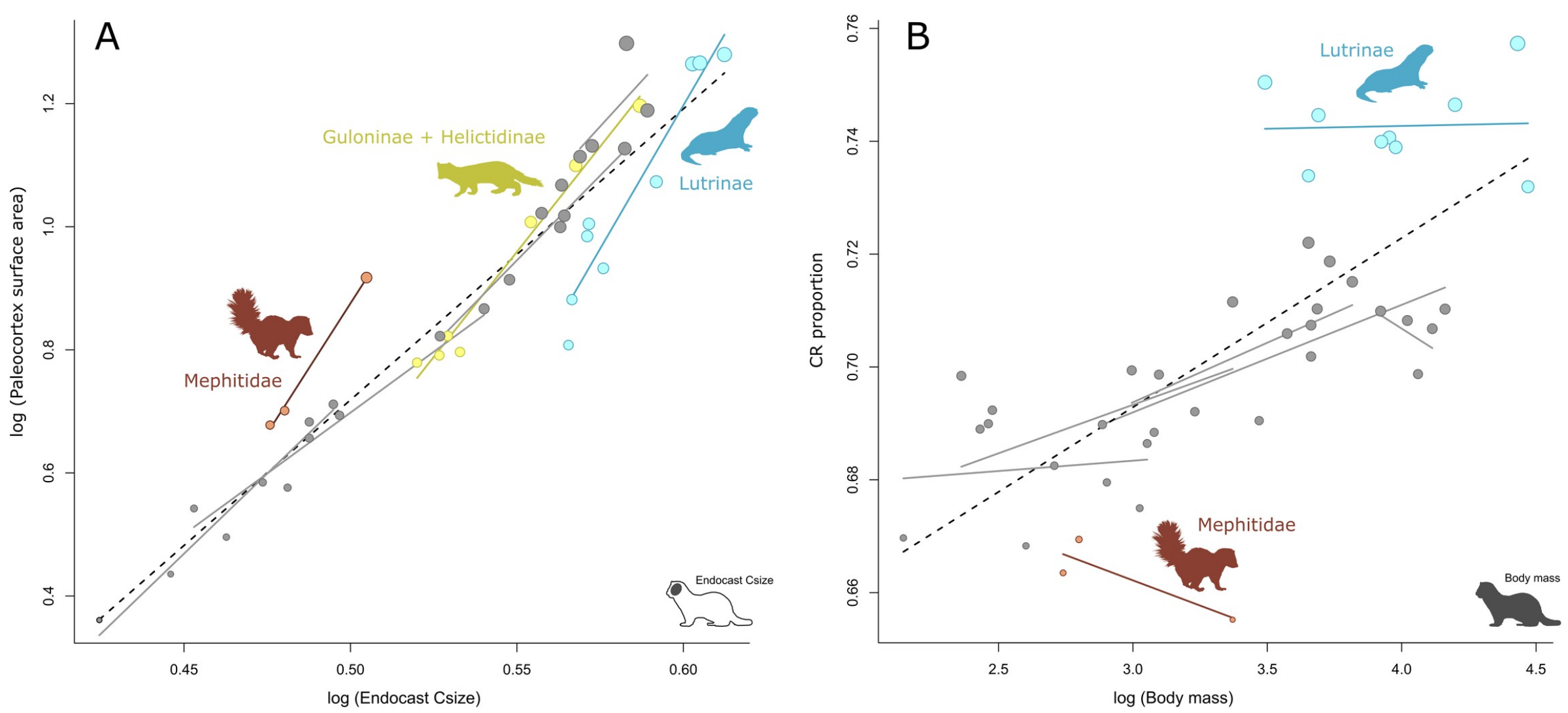

**Figure 8. Differences in allometric scaling across musteloid subfamilies.** (**A**) Paleocortex surface area against endocast Csize, and (**B**) CR proportions against body mass. Point size is scaled to the log-transformed endocast centroid size. Only groups with strong Bayesian support for a deviation from the global allometric slope are shown in color.

# DISCUSSION

## Size as the primary driver of endocranial variation

The results of our comparative analyses show that size consistently emerged as one of the strongest predictors of endocranial variation in Carnivora and is often retained as the best explanatory variable for metrics describing brain organization (notably for ILR z2, hindbrain proportion, paleocortex and neocortex surface areas). Similarly, the optimal PGLS models systematically included size as the sole predictor for shape analyses conducted on the whole brain and its major anatomical regions. However, for both whole brain and regional shape, size explained only a small proportion of total variance (~5–8%, depending on the size metric and region), indicating a consistent but relatively weak allometric constraints on neuroanatomical form. In contrast, size was a major determinant of relative surface areas: for both paleocortex and neocortex evolution, size alone accounted for the vast majority of variance (83–98% across analyses). Robustness analyses showed that backward stepwise models consistently fell within the top-ranked candidate models (ΔAICc < 2), confirming that the dominant role of size is robust to the model-selection strategy.

In agreement with broader mammalian patterns (Finlay & Darlington, 1995; Herculano-Houzel et al., 2014), this provides further support that brain organization in musteloids is mostly structured by allometry. Endocranial proportions follow a clear directional allometric pattern as well: the cerebrum expands proportionally with brain size, whereas the olfactory bulbs and hindbrain decrease in relative volume. In support of this, an expansion of both the paleocortex and the neocortex is also observed as size metrics increase, with the former exhibiting a slightly faster rate of growth than the latter. This scaling relationship is consistent with encephalization pathways seen in many other mammalian clades (Weisbecker et al., 2021; Lang et al., 2025), in which cerebral structures such as the olfactory bulbs grow more slowly and appear proportionately reduced, while others, as the neocortex, are associated with preferential investment when brain size increases (Jerison, 1973; Radinsky, 1977; Reep et al., 2007; Bertrand et al., 2022). Moreover, the relationship between olfactory bulbs proportion and the anterior elongation of the endocast in musteloids lines up with patterns observed in other mammalian clades. For instance, in Euarchontoglires, species with smaller brains usually exhibit proportionally larger olfactory bulbs and more anteriorly elongated endocasts (Lang et al., 2024). Similar allometric scaling has been observed in carnivorans using both endocranial and volumetric datasets (Reep et al., 2007; Sakai et al., 2016). This recurring pattern supports the idea that the sizes of different brain regions are primarily dictated by shared developmental constraints instead of independent selective pressures (Finlay et al., 2001; Barton & Harvey, 2000). Interestingly, while the olfactory bulbs tend to decrease in relative size as brain size and body mass increase, the paleocortex (also involved in olfaction) shows a strong positive allometry. Rather than a simple uniform scaling of structures this pattern may suggest that cortical information-processing circuits may be more strongly conserved than the primary olfactory functions carried out by the bulbs. This allometric difference may also reflect shifts in internal organization and a redistribution of neural resources, two hypotheses that are not mutually exclusive.

Contrary to our initial hypothesis, locomotor mode provides little additional explanatory power beyond size after accounting for shared evolutionary history in explaining variation in the hindbrain partition and external endocast shape proxies analyzed here. This conserved structure decreases proportionately as brain volume and body mass increase, following a typical mammalian allometric pattern. Instead of a true reduction of the hindbrain, this pattern could result from the disproportionate scaling of anterior regions, which grow more quickly and reduce the relative contribution of posterior structures even though they continue to grow in absolute terms (Barton & Harvey, 2000; Smaers et al., 2021). Consistently, cerebellar volume (which is part of the hindbrain) exhibits negative allometry in relation to total brain volume across mammals (Heuer et al., 2023), even if some clade-specific deviations can be observed. For example, cerebellar size in cetaceans can be either enlarged or reduced relative to overall brain size, which has been interpreted as reflecting functional specializations (Müller & Montgomery, 2019). In a similar fashion, the cerebellum evolved at accelerated rates in great apes, producing grade shifts consistent with increases in motor and cognitive complexity (Barton & Venditti, 2014). Yet, even in exceptional cases of lineage-specific cerebellar expansion, the overall hindbrain proportion remains somewhat constant. In addition, key hindbrain circuits, including the connections between vestibular organs and the cerebellum, exhibit significant conservation across vertebrate lineages (Lipovšek & Wingate, 2019). This supports the idea that the hindbrain is a structurally and functionally preserved region of the brain, mainly due to pervasive developmental constraints, and is less labile than other areas of the brain as also illustrated by the remarkably stable cerebellar proportion observed across mammalian orders (Clark et al., 2001).

Another possible explanation for this pattern is that, even though our analyses did not show clear volumetric differences between ecological or taxonomic categories, there may still be functional divergences in hindbrain organizations that do not appear as changes in overall volume. Rather, they might manifest through internal changes that our methodology fails to capture. For instance, its allometric reduction in relative volume may be functionally compensated by increased cerebellar folding, which enhances cortical surface area and processing capacity without a proportional rise in total volume (Heuer et al., 2023). Similarly, cellular analyses have highlighted that the number of cerebellar neurons increases almost at the same rate as cortical neurons, even though the volume of the cerebellum does not increase as rapidly as the size of the whole brain (Herculano-Houzel et al., 2014). In other words, the cerebellum exhibits a higher neuronal density and gains neurons more rapidly than other brain structures, despite its slower volumetric growth. In contrast, the pons and medulla (both components of the hindbrain) are generally found to exhibit lower neuronal densities as their volume increases (Herculano-Houzel et al., 2014). Recent comparative studies reveal that the evolution of large brains predominantly implies changes in neuronal composition (specifically, reduced neuronal density, larger neurons, and elongated axons) rather than proportional expansion of all brain regions (Kverková et al., 2020; Herculano-Houzel, 2022). Taken together, these results suggest that functional adaptations of the hindbrain may occur through internal reorganization instead of quantifiable volumetric expansion. However, it is important to note that we did not measure the volumes of specific hindbrain components, such as the cerebellum and the petrosal lobules, which may exhibit different evolutionary trends.

## Musteloid brain shape

We found no evidence that ecological variables influence overall or regional endocast shape. After accounting for shared evolutionary history, model selection confirmed the predominant influence of size, with the simplest models (i.e., including only endocranial size and phylogenetic non-independence) providing the best fit for several regions and consistently outperforming more complex ecological models. Exhaustive procD.pgls comparisons across the same additive candidate set yielded broadly similar results, reinforcing the absence of detectable ecological effects on endocast shape. Variation in endocast shape corresponds to changes in brain proportions: species with smaller endocasts exhibit a more elongated brain on the cranio-caudal axis, with prominent and rostrally oriented olfactory bulbs, whereas those with larger endocasts show a proportionally expanded and rounded cerebrum. This implies that endocranial geometry in musteloids primarily reflects size-related allometry and lineage-specific evolutionary history. Nevertheless, literature pertaining to other vertebrate taxa indicates that ecological factors can influence brain morphology in specific contexts. For instance, variation in locomotor behavior has been linked to cerebral modifications involved in motor control and spatial processing in several lineages. Species inhabiting distinct locomotor niches exhibit consistent variations in endocast shape, whether among squamates (Macrì et al., 2019), anurans (Manzano et al., 2017), or mammals (Bertrand et al., 2019). Even in birds, where brain shape is highly conserved (Ksepka et al., 2020), ecological variations are correlated with modifications in brain shape regardless of size, suggesting complex niche influences on the anatomy of the brain (Eliason et al., 2021).

In procyonid carnivorans, Ahrens (2014) found that endocranial shape reflects family-level relationships and ecology. Arboreal procyonids exhibit relatively enlarged brain regions involved in motor control, spatial perception, and balance. This finding is noteworthy because our dataset includes the same species sampled by Ahrens (2014), except for *Bassaricyon alleni* and *Nasuella olivacea*. It is possible that these arboreality-related changes reflect lineage-specific evolutionary patterns and are therefore expressed primarily at the family level,, rather than occurring in a consistent manner across all carnivoran families. This could explain why we do not detect clear ecological patterns at broader taxonomic scales in our analysis. Supporting this potential phylogenetic constraint, we also observe a strong phylogenetic influence on overall brain shape and across all major brain regions, consistent with these findings. A further important consideration is that we did not investigate finer neuroanatomical structures possibly associated with arboreal adaptation, such as paraflocculi (i.e., petrosal lobules) size, which have been demonstrated to be associated with locomotor ecology in both carnivorans (Ahrens, 2014) and rodents (Bertrand et al., 2017; 2018; 2021). It is possible that such features represent a more general neuroanatomical correlate of arboreal ecology than the external morphology of the brain itself. However, it should be emphasized that not all studies support a clear link between ecology and endocranial shape. Notably, Weisbecker et al. (2021) also reported no significant effect of ecological factors (including locomotor categories) on endocast shape in marsupials. Their analyses uncovered significant morphological disparities even among closely related species, further illustrating that endocranial shape is a poor predictor of internal brain organization. This body of evidence suggests that ecological specialization probably impacts brain morphology at a macroevolutionary level. However, the outcomes of these effects are not uniform across taxa and may be likely affected by lineage-specific evolutionary and developmental constraints.

Moreover, although whole-brain shape and most regional measures (apart from the olfactory bulbs) showed a significant interaction between size and subfamily, we did not identify any statistically supported allometric differences among subfamilies. Prior research demonstrates that endocranial shape displays differential scaling relative to size across mammalian taxa, highlighting phylogenetic variability in allometric pathways (Sansalone et al., 2020; De Miguel et al., 2022; Lang et al., 2024). It has also been observed that ecological factors can further affect regional patterns, as evidenced in squamates where cerebellar allometry fluctuates with locomotor ecology (Macrì et al., 2019). However, our results tell a different story where subfamily variation primarily reflects intercept shifts rather than genuine differences in allometric scaling, suggesting that musteloid subfamilies differ in their baseline endocast shape but share a common allometric trajectory. Importantly, these interactions were significant only when brain size was used as the scaling variable, but not when using body mass. This pattern suggests that variation in brain shape and major brain regions is driven primarily by intracranial allometric scaling, rather than by overall somatic growth. In other words, neuroanatomical shape diversification in musteloids appears to be more closely linked to brain-specific evolutionary patterns than to general changes in body size.

## Lineage-specific differences

Although size was a major predictor for many of the traits examined, we found that some features of musteloid brain organization are unexpectedly unrelated to size. Notably, the best explanatory models for ILR axis 1 (i.e., balance between olfactory bulbs and the rest of the brain) as well as for the olfactory bulbs (OB) and the cerebrum (CR) proportion did not include size at all. Instead, even after accounting for phylogenetic non-independence, variation in hindbrain proportion was best explained by taxonomic differences among subfamilies, which alone accounted for 43–50% of the variance, revealing surprisingly strong lineage-specific patterns of brain organization within musteloids.

Otters (i.e., subfamily Lutrinae within Mustelidae) emerge as one of the most divergent lineages in our analyses, displaying a significantly distinctive neuroanatomical profile with a proportionally enlarged cerebrum (CR) and reduced olfactory bulbs (OB) relative to other musteloids. Regarding the cerebrum, neuroanatomical investigations of semi-aquatic and aquatic carnivorans have revealed an uneven enlargement of somatosensory cortical areas, particularly the coronal gyrus, linked to tactile foraging and precise forelimb manipulation in aquatic environments (Radinsky, 1968; Marshall et al., 2014; Sawyer et al., 2016). Similar somatosensory adaptations have been postulated in early aquatic carnivoran lineages (Lyras et al., 2023), indicating that this transition towards tactile processing is a persistent feature of aquatic evolution in Carnivora. This may imply that otters possess a cerebrum that is not only larger compared to most other musteloids but may also operate more effectively to process tactile and sensorimotor information. An additional noteworthy pattern is that otters (Lutrinae) display significantly reduced olfactory bulb size compared to other musteloid subfamilies. This pattern is consistent with the results of Gittleman (1991), who found that otters have significantly smaller olfactory bulbs in comparison to other carnivores, implying a reduced dependence on olfaction in aquatic habitats. Skeletal evidence corroborates the trend of olfactory reduction in semi-aquatic species. Indeed, Bird et al. (2014) found that the surface area of the cribriform plate (i.e., the contact area between the olfactory bulbs and the nasal cavity) scales with negative allometry relative to body size and is significantly reduced in aquatic carnivorans. Incidentally, this finding also supports the idea that the smaller olfactory bulbs observed in semi-aquatic Musteloidea are a result of both allometric scaling and a reduced dependence on olfaction within this group.

In addition, our Bayesian hierarchical models reveal detectable lineage-specific departures from the global allometric pattern, consistent with divergent neuroanatomical evolution within the clade. Specifically, Lutrinae and Mephitidae show significantly steeper paleocortex scaling when size is expressed as brain volume, and both lineages also deviate from the global slope for the cerebrum (CR) proportion when size is expressed as endocranial size. These results indicate that some portions of the brain have followed distinct evolutionary trajectories in these groups, even though the general allometric direction is conserved across Musteloidea. These deviations reflect shifts in allometric slopes rather than simple differences in intercept, suggesting that the proportional relationships among brain regions may vary with size along lineage-specific trajectories. This pattern is consistent with the idea that evolutionary change often proceeds along major allometric axes (i.e., line of least evolutionary resistance) rather than through extensive reshaping of brain organization (Marroig & Cheverud, 2005). Nevertheless, these signals should be interpreted cautiously, particularly for Mephitidae, as the small number of species available in this subfamily may amplify stochastic variation and limit the strength of the evolutionary inference.

## Ecological patterns of brain organization

Contrary to our initial prediction, we did not identify a distinct endocranial compositional signature in semi-fossorial species compared to other locomotor categories. More broadly, none of the best-supported models for any of our neuroanatomical variables provided evidence for a significant effect of locomotor mode. This hypothesis was initially supported by prior evidence suggesting that fossorial and subterranean mammals have relatively enlarged olfactory bulbs, which reflects their reliance on olfaction in low-visibility environments (Ribeiro et al., 2014; Tambusso & Fariña, 2015; Napoli et al., 2017; Ruiz-Rubio et al., 2024), although this pattern does not hold across all fossorial mammals, with Aplodontiidae showing a reduction in olfactory structures (Bertrand et al., 2018). Substantial neuroanatomical modifications in the cerebrum have been thoroughly recorded in fossorial mammals. Subterranean lifestyle in taxa such as talpids, bathyergid mole-rats, and spalacids (Catania & Kaas, 1997; Nevo, 1995; Catania, 2000; Bennett & Faulkes, 2000) is associated with substantial cortical reallocation, where visual areas are reduced while somatosensory representations expand substantially, consequently occupying a larger portion of the neocortex. For example, more than half of the primary somatosensory cortex in the star-nosed mole is dedicated to nasal-ray input, demonstrating how much this species depends on tactile information (Catania & Kaas, 1997). Bertrand et al. (2018) showed that aplodontiid rodents exhibit a reduced neocortex, indicating that a decrease in neocortical size can arise as a response to fossorial specialization. Nonetheless, these shifts may also occur within the cerebrum and do not inherently modify brain regions volumes. Studies on rodents have demonstrated that cortical specialization can arise through local remapping of the brain, without changes in overall neocortical or forebrain volume (Catania & Remple, 2002; Campi & Krubitzer, 2010; Krubitzer et al., 2011), which is consistent with our results suggesting that size, rather than ecological specialization, is the main determinant of neocortex surface area in musteloids. This implies that adaptations to burrowing might operate largely at a fine-scale cortical level, beneath the resolution achieved by endocast regional partitions. It is therefore possible that semi-fossorial musteloids exhibit comparable, subtle neural modifications facilitating subterranean lifestyle, but lack any apparent changes in gross endocranial composition. In addition, the association between fossorial habits and enlarged olfactory bulbs appears to be clade-dependent (Ribeiro et al., 2014). Finally, a more likely scenario is that such changes are minor or even absent, since semi-fossorial musteloids are not obligatory subterranean specialists. In fact, many species use burrows primarily for shelter or nesting while still foraging and navigating above ground (Noonan et al., 2015). As such, they do not experience the extreme sensory channel narrowing typical of fully subterranean mammals, which likely weakens selective pressures for strong cortical or volumetric reorganization. It is important to note that a decrease in the size of one cerebellar region, the petrosal lobules, was observed in aplodontiid rodents (Bertrand et al., 2018), fossorial Euarchontoglires (Lang et al., 2022) and subterranean mammals more generally (Goyens et al., 2022). This brain region was, however, not quantified here. The petrosal lobules have a role in maintaining the eyes and head during locomotion (Rambold et al., 2002; Hiramatsu et al., 2008) and these functions might be less crucial for survival in a subterranean environment.

Another notable finding is that, in musteloids, the balance between anterior (OB + CR) and posterior (HB) brain regions is shaped by both overall size and diet. Moreover, when the analyses are performed at the family level, a different pattern emerges with diet emerging as the primary predictor of olfactory bulb (OB) proportion. In particular, we detect clear differences between piscivorous species and those that are carnivorous or omnivorous. Although these patterns should be interpreted with caution given their partial dependence on the modelling framework, complementary robustness analyses recovered broadly similar results, suggesting that both the detected ecological patterns and the absence of locomotor effects are not artefacts of the specific model-selection strategy used. These results demonstrate that different selective pressures operate at different evolutionary scales and highlight the significant role of feeding ecology in driving neuroanatomical diversification within musteloids. One possible explanation is that feeding-related mechanical constraints shape the cranial architecture surrounding the brain. The anatomy of the masticatory muscle's places distinct spatial and developmental constraints on the cranium, depending on the size and arrangement of the individual muscles. These constraints are even more important at the level of muscle insertion areas where mechanical loading influences bone growth, shape, and orientation. (Christiansen & Adolfssen, 2005; Christiansen & Wroe, 2007; Dickinson et al., 2021; Siciliano-Martina et al., 2021; Ito et al., 2024). The observed differences between piscivorous and their other relatives may be explained by these interactions between bone structure, jaw muscles, and brain development. It supports previous research showing that adaptation to feeding ecology is reflected in musteloids cranial morphology variation (Dumont et al., 2016; Law et al., 2018a), a pattern noticed in other aquatic carnivorans, most of which being predominantly piscivorous (Jones & Goswami, 2012; Kienle & Berta, 2016, 2018). It is also worth noting that most piscivorous species in our dataset are Lutrinae, and that similar intraspecific correlations among cranial morphology, musculature, and feeding ecology have been recorded in otters (Russo et al., 2022, but cf. Law et all., 2018a). Thus, this apparent ecological effect might be the result of a combination of shared evolutionary history and dietary influences. To gain a more comprehensive understanding of the functional and developmental interactions shaping mammalian brain evolution, future research should integrate endocranial, cranial, and muscular data within a broader evolutionary context with several ecological convergences events to better distinguish among these effects.

## Conclusion

Combining information about the shape and composition of the endocranium reveals a coherent view of brain evolution in musteloids: a neuroanatomical organization that depends primarily on size-dependent scaling, with coordinated shifts in both regional proportions and endocast geometry as brains enlarge. Some lineages nevertheless differ markedly from this conserved pattern. Otters (Lutrinae) and skunks (Mephitidae) display distinctive proportional arrangements of cortical and olfactory regions, underscoring how lineage history influences brain composition. Feeding ecology also contributes to this diversity, with piscivorous species exhibiting a characteristic anterior–posterior balance linked to trophic specialization and its anatomical constraints. Overall, musteloid brain diversity arises from the interplay between strong allometric constraints and more localized lineage- and ecology-driven shifts. Future comparative work integrating endocranial, cranial, and behavioral data across multiple independent ecologies will be essential to test the generality of these findings.

# AUTHOR CONTRIBUTIONS

MM conceived the study and prepared the initial draft of the manuscript. ARH, SL and MM acquired the CT scans files. MM and AA conducted the segmentation. EGL and OB contributed to the refinement and optimization of the study methodology. All authors critically reviewed the manuscript and approved the final version.

# ACKNOWLEDGEMENTS

We thank the Fyssen Foundation for supporting this research through a fellowship during which part of the specimen sampling was carried out. Scanning of specimens by ARH was funded by NSF BCS 1825129. We thank Davide Tamagnini for providing three endocasts (*Amblonyx cinereus* AMNH 101638; *Martes melampus* PRIZ475; *Martes melampus* PRIZ572). We also thank Adrien de Perthuis for segmenting some 3D virtual endocasts (*Lontra canadensis* UF 23593). We are grateful to Emmanuel Gilissen for providing specimens as part of the Fyssen Foundation project (RMCA). We thank the curators and staff of the following museums for the loan of specimens: G. Véron & C. Bens (MNHN); L. Gatens & B. Hess (NCSM); V. Mathis & D. Reed (UF); M. Surovy, S. Ketelsen, E. Hoeger, & N. Duncan (AMNH). We thank the Digital Morphology Museum KUPRI and the Primate Research Institute of Kyoto University (PRIZ). We thank all contributors and managers involved in creating, uploading, and curating the specimens hosted on MorphoSource: J. Tseng (AMNH); O. Pauwels (RBINS); B. Van Valkenburgh, T. Rowe, B. Sulbaran, E. Cote & J. Maisano (USNM); G. Pandelis & C. Bick (UMMZ); L. Lamb & C. Yoakum (UAMC); B. Van Valkenburgh, T. Rowe, J. Maisano, E. Cote & M. Colbert (UCLA). The institutional affiliations indicated above correspond to those associated with the contributors on MorphoSource at the time the datasets were uploaded. The specimen of *Lontra felina* (AMNH M-48193) belongs to a collection that was funded by the NSF, the Yunnan Natural Science Foundation, the Government of Zhaotong, the IVPP, the NSFC, and the governments of Zhaotong and Zhaoyang. The specimens *Mephitis mephitis* (USNM 147553), *Neogale frenata* (USNM 95054), *Potos flavus* (USNM 291066), and *Taxidea taxus* (UCLA 14841) belong to a collection whose data acquisition was funded by NSF IOB-0517748, and data upload to MorphoSource was funded by DBI-1902242. Finally, we thank the reviewers for their constructive comments and suggestions which helped improve the clarity and robustness of this manuscript. M.M. also thanks Carlo Meloro for valuable discussions and advice regarding the interpretation and terminology of phylogenetic signal.

# CONFLICT OF INTEREST STATEMENT

The authors declare no conflicts of interest.

# DATA AVAILABILITY STATEMENT

The data that support the findings of this study are available in the Supporting Information of this article.

# ORCID

Margot MICHAUD: 0000-0001-7075-9862
Arianna HARRINGTON: 0000-0002-2720-6098
Sandrine LADEVEZE: 0000-0001-6009-4107
Eloy GÁLVEZ LÓPEZ : 0000-0002-6211-966X
Valentin FISCHER : 0000-0002-8808-6747
Apolline ALFSEN: 0009-0002-8678-3888
Ornella BERTRAND : 0000-0003-3461-3908